\documentclass[10pt,aps,physrev,reprint,floatfix,superscriptaddress,longbibliography]{revtex4-2}
\pdfoutput=1
\usepackage{graphicx,latexsym}
\usepackage{dcolumn,amsfonts}
\usepackage{amssymb,amsmath,bm}

\usepackage[breaklinks=true,pdfencoding=auto]{hyperref}

\usepackage{natbib}
\usepackage{balance}

\hypersetup{
	colorlinks   = true, 
	urlcolor     = blue, 
	linkcolor    = blue, 
	citecolor    = red 
}

\usepackage{color}
\usepackage{ulem}

\begin{document}

	\title{Controlling the excitation spectrum of a quantum dot array
		   with a photon cavity}

	\author{Vidar Gudmundsson}
	\email{vidar@hi.is}
	\affiliation{Science Institute, University of Iceland, Dunhaga 3, IS-107 Reykjavik, Iceland}
	\author{Vram Mughnetsyan}
	\email{vram@ysu.am}
	\affiliation{Department of Solid State Physics, Yerevan State University, Alex Manoogian 1, 0025 Yerevan, Armenia}
	\author{Nzar Rauf Abdullah}
	\affiliation{Physics Department, College of Science,
		University of Sulaimani, Kurdistan Region, Iraq}
	\affiliation{Computer Engineering Department, College of Engineering, Komar University
		of Science and Technology, Sulaimani 46001, Kurdistan Region, Iraq}
	\author{Chi-Shung Tang}
	\email{cstang@nuu.edu.tw}
	\affiliation{Department of Mechanical Engineering, National United University, Miaoli 36003, Taiwan}
	\author{Valeriu Moldoveanu}
	\email{valim@infim.ro}
	\affiliation{National Institute of Materials Physics, PO Box MG-7, Bucharest-Magurele,
		Romania}
	\author{Andrei Manolescu}
	\email{manoles@ru.is}
	\affiliation{Department of Engineering, Reykjavik University, Menntavegur
		1, IS-102 Reykjavik, Iceland}

%

\begin{abstract}
We use a recently proposed quantum electrodynamical density functional theory (QEDFT)
functional in a real-time excitation calculation for a two-dimensional electron gas
in a square array of quantum dots in an external constant perpendicular magnetic field
to model the influence of cavity photons on the excitation spectra of the system.
The excitation is generated by a short elecrical pulse. The quantum dot array is defined
in an AlGaAs-GaAs heterostructure, which is in turn embedded in a parallel plate
far-infrared photon-microcavity. The required exchange and correlation energy functionals
describing the electron-electron and electron-photon interactions have therefore been
adapted for a two-dimensional electron gas in a homogeneous external magnetic field.
We predict that the energies of the excitation modes activated by the pulse are generally
red-shifted to lower values in the presence of a cavity. The red-shift can be understood in terms
of the polarization of the electron charge by the cavity photons and depends on the magnetic flux,
the number of electrons in a unit cell of the lattice, and the electron-photon interaction strength.
We find an interesting interplay of the exchange forces in a spin polarized two-dimensional
electron gas and the square lattice structure leading to a small but clear blue-shift of
the excitation mode spectra when one electron resides in each dot.
\end{abstract}

\maketitle
%
%

\section{Introduction}
The interest in using photon cavities
to control and influence processes in chemistry and properties of
physical systems has been growing in recent years
\cite{https://doi.org/10.1002/anie.201107033,Flick2017,FlickRiveraNarang+2018+1479+1501,doi:10.1021/acs.nanolett.9b02982,10.1063/1.5142502,PhysRevA.106.053710}.
At the same time experiments show that a two-dimensional electron gas
(2DEG) in an AlGaAs-GaAs heterostructure can be placed in a
high-quality-factor terahertz cavity in an external magnetic field,
and the high polarizability of the 2DEG makes it possible to
attain a non-perturbative coupling of electrons with the
cavity photons \cite{Zhang1005:2016}.

The influence of the electron-photon interaction on the properties of a
homogeneous free 2DEG (without the mutual Coulomb interaction between the electrons) in a parallel
plate photon microcavity and an external magnetic field have been studied
\cite{PhysRevResearch.4.013012,PhysRevLett.123.047202,PhysRevB.105.205424}.

The effects of the electron-photon interaction together with the electron-electron
Coulomb interaction have been modeled in
nanoscale electronic systems with few electrons using configuration
interaction (CI) approaches (exact numerical diagonalization) in truncated
many-body Fock-spaces \cite{PhysRevE.86.046701,Jonasson2011:01,ANDP:ANDP201500298}.
For electron systems with a higher number of electrons different quantum electrodynamical
density functional approaches have been used, involving coupling of the Maxwell
equations through response functions \cite{doi:10.1073/pnas.1518224112,Buchholz2019,Rubio2021:2110464118},
or the use of explicit QEDFT functionals with no direct photon variables \cite{flick2021simple},
and direct inclusion of electron-photon coupling into density functional theory (DFT) \cite{10.1063/5.0123909}.

An excellent introduction to, and a review on, time-dependent density functional theory (TDDFT)
can be found in the book of Carsten A.\ Ullrich \cite{CA-Ullrich:Book2012}.
In our previous work \cite{PhysRevB.106.115308} we revealed the effects of electron-photon interaction
on the orbital and spin magnetization of a cavity-embedded quantum dot array. Here we test the QEDFT approach
in a non-equilibrium setting and investigate the excitation of collective modes (e.g.\ dipole and quadrupole density modes) by a real-time pulse. Different types of real-time approaches have been applied to electronic systems described with DFT models
\cite{Lopata2011,Tandiana2021,https://doi.org/10.1002/wcms.1341,Trepl2022,doi:10.1021/acs.jctc.8b00580,10.1063/5.0123909}, just to cite few.
In order not to be bound to the linear response regime, we rely on the Liouville-von Neumann equation (L-vNE)
for the density operator \cite{PhysRevB.105.155302}. This approach to the time-evolution of the system
after excitation was originally employed by the present group for investigating a spinless Hartree-interacting
periodic 2DEG in an external magnetic field \cite{PhysRevB.105.155302}.
To this formalism we make modifications in order to incorporate the spin degree of freedom, the exchange
and correlation potentials and the corresponding functional for the electron-photon interaction.

Encouraging to our real-time QEDFT modeling is our former investigation
that found a red-shift of the dipole excitation modes in a model of
few electrons interacting with one cavity photon mode, where the
electron-electron and the electron-photon interactions are described
with an exact numerical diagonalization in a truncated Fock-space, i.e.\
where a configuration interaction is used \cite{ANDP:ANDP201500298}.
In Ref.\ \onlinecite{ANDP:ANDP201500298}, and later publications \cite{doi:10.1002/andp.201900306},
where a full account is taken of the electron confinement potential and the shape of the resulting
charge density for a double-dot system, we studied electron transport through an open
electron-photon system, but here, in an extended two-dimensional array of quantum dots
the exact diagonalization method becomes intractable.

The paper is organized as follows:
In Section \ref{Model} we describe first briefly the static model in Subsection
\ref{StaticModel}, and subsequently the modeling of the time-evolution of the excited
system in Subsection \ref{StaticModel}. The results and discussion thereof are found in
Section \ref{Results}, with the conclusions drawn in Section \ref{Conclusions}.

\section{Model}
\label{Model}
We consider an array of interacting quantum dots defined in a 2DEG, under external excitation,
and, additionally, placed in a cavity which supports a single photonic mode of energy $\hbar\omega_{\gamma}$. We assume that the 2DEG is placed in the
centered $x-y$ plane of the cavity. To model this complex system we follow a two-step strategy:
i) First, we assume that at equilibrium the hybrid cavity-QD array device is described by
an effective single-electron static Hamiltonian $H_{{\rm stat}}$
which embodies the electron-electron and electron-photon interactions via suitable
exchange-correlation functionals;
ii) Secondly, the effects of the real-time excitation $V^{{\rm ext}}$ are analyzed by solving
the Liouville-von Neumann equation associated to $H_{{\rm stat}}+V^{{\rm ext}}({\bf r},t)$.
Let us recall here that this approach allows us to go beyond the linear response approximation
(see Ref.\ \onlinecite{PhysRevB.105.155302}). Note also that in the present work
the application of the L-vNE is extended by including the spin degree of freedom and
the exchange-correlation Coulomb functionals.

\subsection{The static system}
\label{StaticModel}
The dynamically two-dimensional electrons reside in a square lattice
of quantum dots defined by the potential
\begin{equation}
	V_\mathrm{per}(\bm{r}) = -V_0\left[\sin \left(\frac{g_1x}{2} \right)
	\sin\left(\frac{g_2y}{2}\right) \right]^2,
	\label{Vper}
\end{equation}
where $V_0 = 16.0$ meV.
\begin{figure}[htb]
	\includegraphics[width=0.44\textwidth]{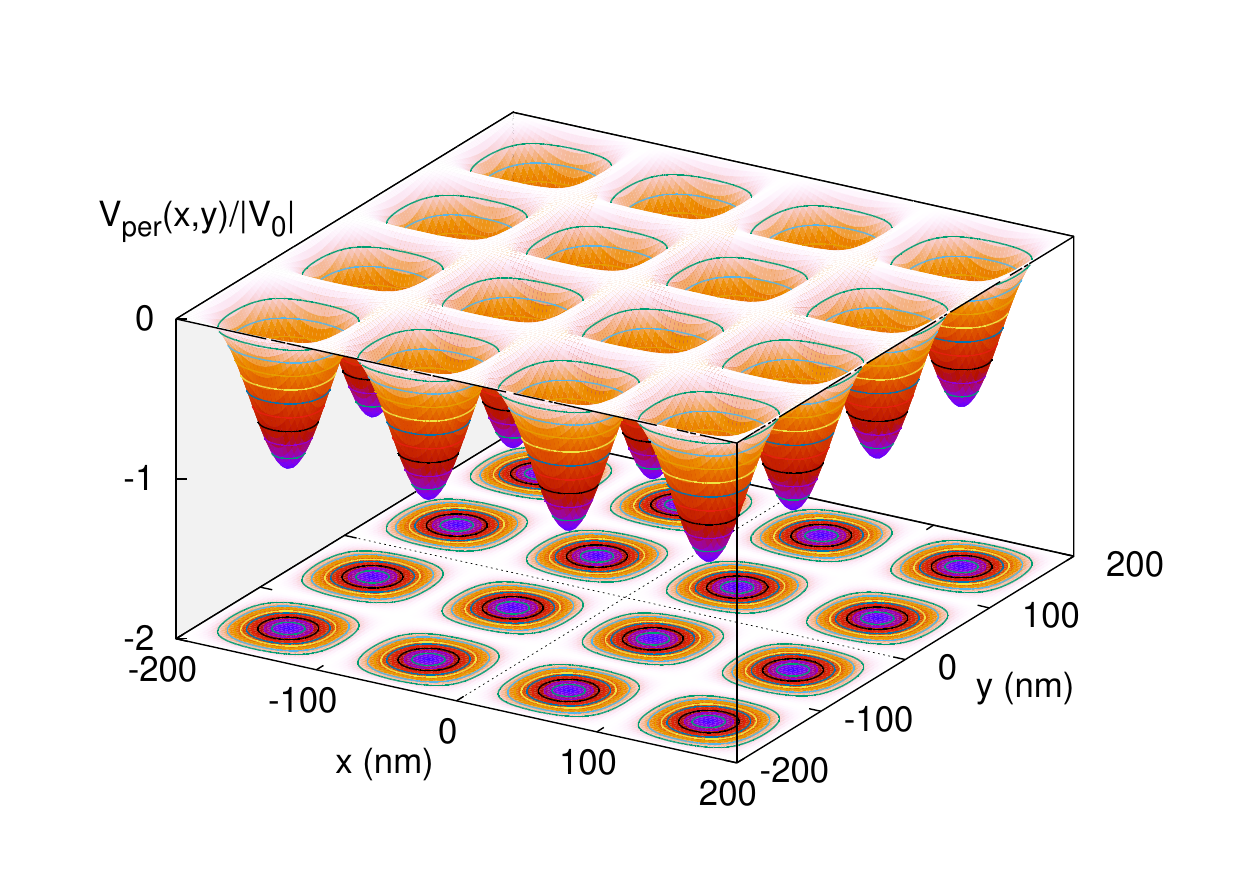}
	\caption{The periodic potential $V_\mathrm{per}(x,y)$ defining the square
		array of quantum dots.}
	\label{Vxy2}
\end{figure}
The superlattice array is defined by the vectors
$\bm{R}=n\bm{l}_1+m\bm{l}_2$, where $n,m\in \bm{Z}$. The unit vectors
are $\bm{l}_1 = L\bm{e}_x$ and $\bm{l}_2 = L\bm{e}_y$.
The inverse lattice is formed by $\bm{G} = G_1\bm{g}_1 + G_2\bm{g}_2$ with
$G_1, G_2\in \bm{Z}$ and the unit vectors
\begin{equation}
	\bm{g}_1 = \frac{2\pi\bm{e}_x}{L}, \quad\mbox{and}\quad
	\bm{g}_2 = \frac{2\pi\bm{e}_y}{L},
\end{equation}
where $L = 100$ nm is the lattice period. The number of electrons in each dot is noted by $N_e$.
We use GaAs parameters, $m^* = 0.067m_e$, $\kappa = 12.4$, and $g^* = 0.44$.
The 2DEG is in a perpendicular homogeneous magnetic field $\mathbf{B} = B\hat{\mathbf{z}}$ expressed
by the vector potential $\mathbf{A} = (B/2)(-y,x)$ using the circular gauge in order to
facilitate analogous handling of the $x$ and $y$ Cartesian coordinates selected here
for represention of the wavefunctions and the electron density.

The interacting many-body system will be described by an effective single-electron
Hamiltonian implied by the Kohn-Sham equations \cite{PhysRev.140.A1133}
\begin{equation}
	H_\mathrm{stat} = H_0 + H_\mathrm{Zee} + V_\mathrm{H} + V_\mathrm{per} + V_\mathrm{xc} + V^\mathrm{EM}_\mathrm{xc},
	\label{Hstat}
\end{equation}
where
\begin{equation}
	H_0 = \frac{1}{2m^*}\bm{\pi}^2, \quad\mbox{with}\quad
	\bm{\pi} = \left(\bm{p}+\frac{e}{c}\bm{A} \right).
	\label{H0}
\end{equation}
$H_\mathrm{Zee} = \pm g^* \mu_\textrm{B}^* B/2$ is the Zeeman term, and the direct Coulomb interaction is given
by the Hartree potential
\begin{equation}
	V_\mathrm{H}(\bm{r}) = \frac{e^2}{\kappa}\int_{\bm{R}^2}d\bm{r}'\frac{\Delta n(\bm{r}')}
	{|\bm{r}-\bm{r}'|},
	\label{Vcoul}
\end{equation}
where $\Delta n(\bm{r}) = n_e(\bm{r})-n_\mathrm{b}$, with $+en_\mathrm{b}$ being the
homogeneous positive background charge density needed to guarantee charge neutrality of the
total system, and $-en_e(\bm{r})$ is its electron charge density.
$\mu_\textrm{B}^*$ is the effective Bohr magneton.
The potential $V_\mathrm{xc}$ represents the exchange and correlation effects due to the
Coulomb interaction within the local spin-density approximation (LSDA) \cite{PhysRevB.106.115308},
and $V^\mathrm{EM}_\mathrm{xc}$ is the corresponding potential derived from the exchange and
correlation energy functional associated with the electron-photon interactions introduced by
Flick \cite{flick2021simple} and adapted to a 2DEG in a magnetic field by the present authors
\cite{PhysRevB.106.115308}. Importantly, this last functional depends on the electron density
and its gradient and represents the first attempt to establish a ``simple'' functional for
a quantum electrodynamic density functional theory (QEDFT)
based on the adiabatic-connection fluctuation-dissipation theorem
\cite{PhysRevB.65.235109,doi:10.1080/00268976.2011.614282}
to describe interacting electron-photon systems \cite{flick2021simple}.
Interestingly, parallels can be found in the description of the van der Waals interaction,
a dynamical interaction of mutually induced dipoles \cite{PhysRevLett.103.063004,PhysRevA.81.062708}.

The spin-dependent exchange-correlation potential $V_\mathrm{xc}$ is calculated as
\begin{equation}\label{V_xc_spin}
      V_{\mathrm{xc},\sigma}(\bm{r},B)=\frac {\partial}{\partial n_{\sigma}}
      (n_e\epsilon_\mathrm{xc}[n_{\uparrow},n_{\downarrow}, B])
      |_{n_{\sigma}=n_{\sigma}(r)}
\end{equation}
using the exchange and correlation functional $\epsilon_{{\rm xc}}$ calculated in
Appendix A of Ref.\ \onlinecite{PhysRevB.106.115308}.
The exchange and correlation potential for the electron-photon interaction is given as
\begin{equation}
      V_\mathrm{xc}^\mathrm{EM} =
      \left\{ \frac{\partial}{\partial n_e} -
      \bm{\nabla}\cdot\frac{\partial}{\partial\bm{\nabla}n_e}\right\}
      E_\mathrm{xc}^\mathrm{GA}(n_e,\bm{\nabla} n_e,\lambda)
\end{equation}
with the electron-photon interaction strength $\lambda$.
More technical details on the exchange and correlation functionals
for the electron-electron and electron-photon interactions are to be found in the
Appendices of Ref.\ \onlinecite{PhysRevB.106.115308}.

In order to maintain an analogous handling of the Cartesian $x$ and $y$ coordinates in
the wavefunctions we use a basis designed by Ferrari \cite{Ferrari90:4598} and used
by others to investigate the properties of a modulated 2DEG in a magnetic field
\cite{Silberbauer92:7355,Gudmundsson95:16744}. In order to use a notation for the
states and wavefunctions in accordance with recent publications
\cite{PhysRevB.105.155302,PhysRevB.106.115308}
we use $|\bm{\alpha}\bm{\theta}\sigma)$ and $\psi_{\bm{\alpha}\bm{\theta}\sigma}(\bm{r})$ for the states
and the wavefunctions (orbitals) of the 'QEDFT interacting' equilibrium 2DEG
(\ref{Hstat}), described in details Ref.\ \onlinecite{PhysRevB.106.115308}. $\bm{\alpha}$ is a composite
quantum number labeling the Landau-bands and their subbands. Each Landau-band is split into $pq$
subbands, where $pq$ is the integer number of magnetic flux quanta flowing through the unit of the
square $L\times L$ 2D Bravais lattice. $\sigma \in \{\uparrow,\downarrow\}$ is the $z$-spin component.
and $\bm{\theta} = (\theta_1,\theta_2) \in \{[-\pi,\pi]\times[-\pi,\pi] \}$ is the location in
the first Brillouin zone (BZ). The external homogeneous magnetic field defines the magnetic length
$l = (\hbar c/(eB))^{1/2}$ and in order to have it commensurate with the lattice length $L$ we set
$2\pi l^2 pq = L^2$ here \cite{Ferrari90:4598,Hofstadter76:2239}.
Due to the commensurate length scales, $L$ and $l$, and the properties of the electron-electron
Coulomb interaction and of the electron-photon interaction the time-independent problem can be solved
in each point of the BZ leading to Landau-band dispersion information, like is shown in Figure 4
in Ref.\ \onlinecite{PhysRevB.106.115308}.

\subsection{Time-evolution of the excited system}
\label{TimeModel}
As we excite the strongly modulated 2DEG with a circularly polarized external potential
pulse
\begin{align}
 	V^\mathrm{ext}(\bm{r},t) = V_\mathrm{t} \left\{ (\Gamma t)^2 e^{-\Gamma t}\right\}
 	[ &\cos{(k_y y)}\cos{(\Omega t)} \nonumber \\
 	+ c_r&\cos{(k_x x)\sin{(\Omega t)}} ]
 	\label{phi-ext}
\end{align}
with $c_r=\pm 1$, a new length scale defined by the wavevector $(k_x,k_y)$
will be forced on the system mixing up states at
different points in the BZ. To accommodate to this in the calculations we need to extend the BZ to
$\bm{\theta} \in \{[-p\pi,p\pi]\times[-q\pi,q\pi] \}$ in order to use the periodicity of $\psi_{\bm{\alpha}\bm{\theta}\sigma}(\bm{r})$ with respect to $\bm{\theta}$ to ensure
the continuity of the wavefunctions across zone boundaries in reciprocal space.

In order to continue with the viewpoint of describing the cavity photon interacting 2DEG system with
an effective single-electron Hamiltonian the total Hamiltonian of the time-dependent system is
\begin{equation}
	H(t) = H_\mathrm{stat} + V^\mathrm{ext}(\bm{r},t)
\label{Ht}
\end{equation}
with $H_\mathrm{stat}$ defined by Eq.\ (\ref{Hstat}). The time-evolution of the density operator
$\rho$ is governed by the nonlinear Liouville-von Neumann equation (L-vNE)
\begin{equation}
	i\hbar\partial_t\rho(t) = \left[ H[\rho (t)],\rho(t) \right] .
\label{L-vNE}
\end{equation}
For the first four terms of $H_\mathrm{stat}$ (on the r.h.s.\ of Eq.\ (\ref{Hstat})) we use the
methods introduced in Ref.\ \onlinecite{PhysRevB.105.155302}, but here we obtain coupled equations
for $\rho_\uparrow$ and $\rho_\downarrow$ in the extended Hilbert space formed by the states
$|\bm{\alpha}\bm{\theta}\sigma)$.

The direct electron Coulomb repulsion term (\ref{Vcoul}) is linear in $\Delta n$, but the
exchange and correlation terms used to derive $V_\mathrm{xc}$ and $V_\mathrm{xc}^\mathrm{EM}$
are generally not linear in $n_e$.
The terms depending on the electron density $n_e(\bm{r})$ are constructed in
$\bm{r}=\bm{x}+\bm{R}$ with $\bm{x}$ in the primitive unit of the square Bravais lattice
and the potentials $V_\mathrm{xc}(\bm{r})$ and $V_\mathrm{xc}^\mathrm{EM}(\bm{r})$ are then
Fourier transformed to reciprocal space $\bm{q}=\bm{k}+\bm{G}$ with $\bm{k}$ in the primitive
inverse lattice. The choice of the impulse, or the wavevector, in the excitation pulse
(\ref{phi-ext}) has to be done in accordance with the truncations for $\bm{R}$ and $\bm{G}$
in the construction of these terms. In order to maximize the accuracy of the components
of the density gradient $\bm{\nabla}n_e$ an additional pair of Fourier transforms is used.

The L-vNE (\ref{L-vNE}) is solved on a time grid and within each time-step
iterations are used to consistently update the spin densities, both density operators,
and the exchange and correlation potentials derived from them \cite{PhysRevB.105.155302}.

To analyze the effects of the excitation and the electron-photon coupling on
the periodic array of quantum dots at each time step, we calculate the averages
$\langle\hat{O} \rangle = \mathrm{Tr}\{\hat{O}\rho (t)\}$
for the dipole operators $\hat{O} = \hat{x}$ and $\hat{O} = \hat{y}$, the quadrupole operator
$\hat{O} = \hat{yx}-\langle\hat{y}\rangle\langle\hat{x}\rangle$, and the monopole operator
$\hat{O} = \hat{x}^2+\hat{y}^2-\langle\hat{x}\rangle^2-\langle\hat{y}\rangle^2$ with
the matrix elements of $\hat{O}$ evaluated with a spatial integral over just one
unit cell \cite{Puente01:235324,PhysRevB.105.155302}.
We label the quadrupole operator as $Q_2$ and use $Q_0$ for
the monopole operator. All the above averages are related
to collective modes with density variations. In order observe rotational collective
divergence free modes (transverse modes) we use the current density
$\bm{j}=-e\dot{\bm{r}} = -(ie/\hbar)[H(t),\bm{r}]$
and calculate \cite{Gudmundsson03:161301}
\begin{equation}
	Q_{\bm{j}} = \frac{1}{l^2\omega_c}
	\langle i(\bm{r}\times\dot{\bm{r}})\cdot\hat{\bm{z}}\rangle ,
	\label{Qj}
\end{equation}
which is directly proportional to the orbital part of the magnetization measured
in one cell, or the orbital angular momentum in the cell. $Q_j$ is connected to the
cyclotron resonance, which is more familiar in homogeneous 2DEGs, but is also
present in modulated systems \cite{Merkt96:1134,PhysRevB.105.155302,PhysRevB.78.245311}.
In addition, for selected points in time we evaluate the time-dependent induced electron
density $\delta n_e(\bm{r},t) = n_e(\bm{r},t) - n_e(\bm{r},0)$. The time-dependent electron
density is evaluated via the density operator
\begin{align}
	n_e(\bm{r},t) =& \mathrm{Tr}\{\delta (\hat{\bm{r}}-\bm{r})\rho (t) \} =\nonumber\\
	\frac{1}{(2\pi)^4}&\int d\bm{\theta}d\bm{\theta}'
	\sum_{\alpha\beta\sigma}\psi^*_{\alpha\bm{\theta}\sigma}(\bm{r})
	\psi_{\beta\bm{\theta}\sigma'}(\bm{r})\rho_{\beta\bm{\theta}',\alpha\bm{\theta}}^\sigma(t).
	\label{Ne}
\end{align}

\section{Results}
\label{Results}
For the initial excitation pulse we use $\hbar\Omega = 6.5$ meV, $\hbar\Gamma = 1.5$ meV,
$V_\mathrm{t} = 0.01$ meV, and $c_r=+1$. The time step is chosen to be $0.02$ ps and the time duration of the pulse
is from $t=0$ to 8 ps. The time-evolution of the system is calculated for 100 ps, unless otherwise
stated for individual runs. The electron-photon interaction strength $\lambda l$ is measured in
$\sqrt{\mathrm{meV}}$ \cite{flick2021simple,PhysRevB.106.115308} and only one photon mode is
considered with energy $\hbar\omega_\gamma = 1.0$ meV. Whether we evaluate the total mean energy
of the system directly as an average of the time-dependent Hamiltonian (\ref{Ht}), as is done
in the Hartree approximation, or take the more appropriate DFT point of view, the total mean energy
remains constant after the initial pulse dies out at $t=8$ ps and the trace of the density
matrix remains constant.

We remind the reader here, that the cavity photons are not present in the calculation, but
their effects on the electron density are included in the electron-photon exchange and correlation
energy functional. The only external parameters to the electron-photon exchange and correlation
energy functional are the coupling $\lambda l$ and the photon energy $\hbar\omega_\gamma$.
Otherwise, it depends only on the electron density and its gradient.

\begin{figure*}[htb]
	\includegraphics[width=0.44\textwidth,bb= 0 5 340 250]{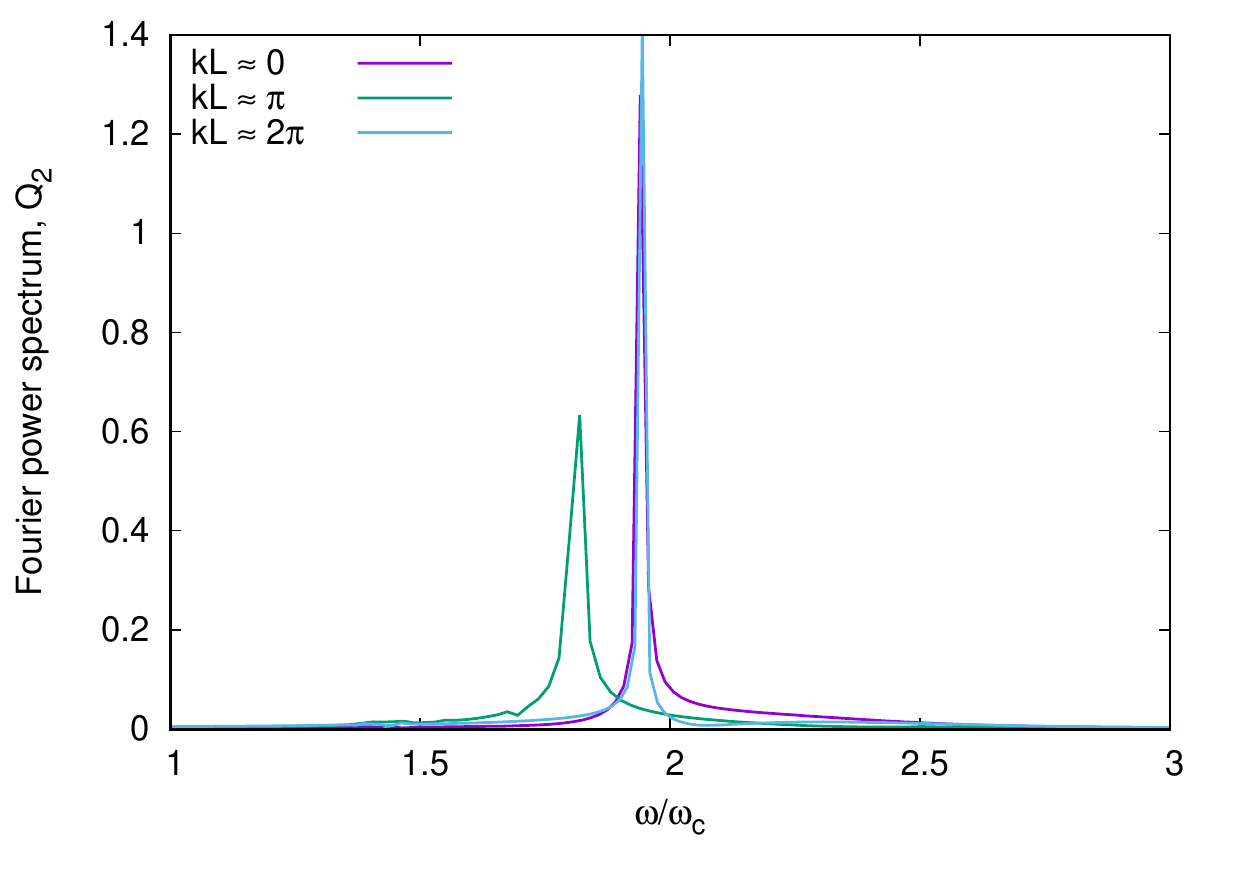}
    \includegraphics[width=0.44\textwidth,bb=25 3 300 280]{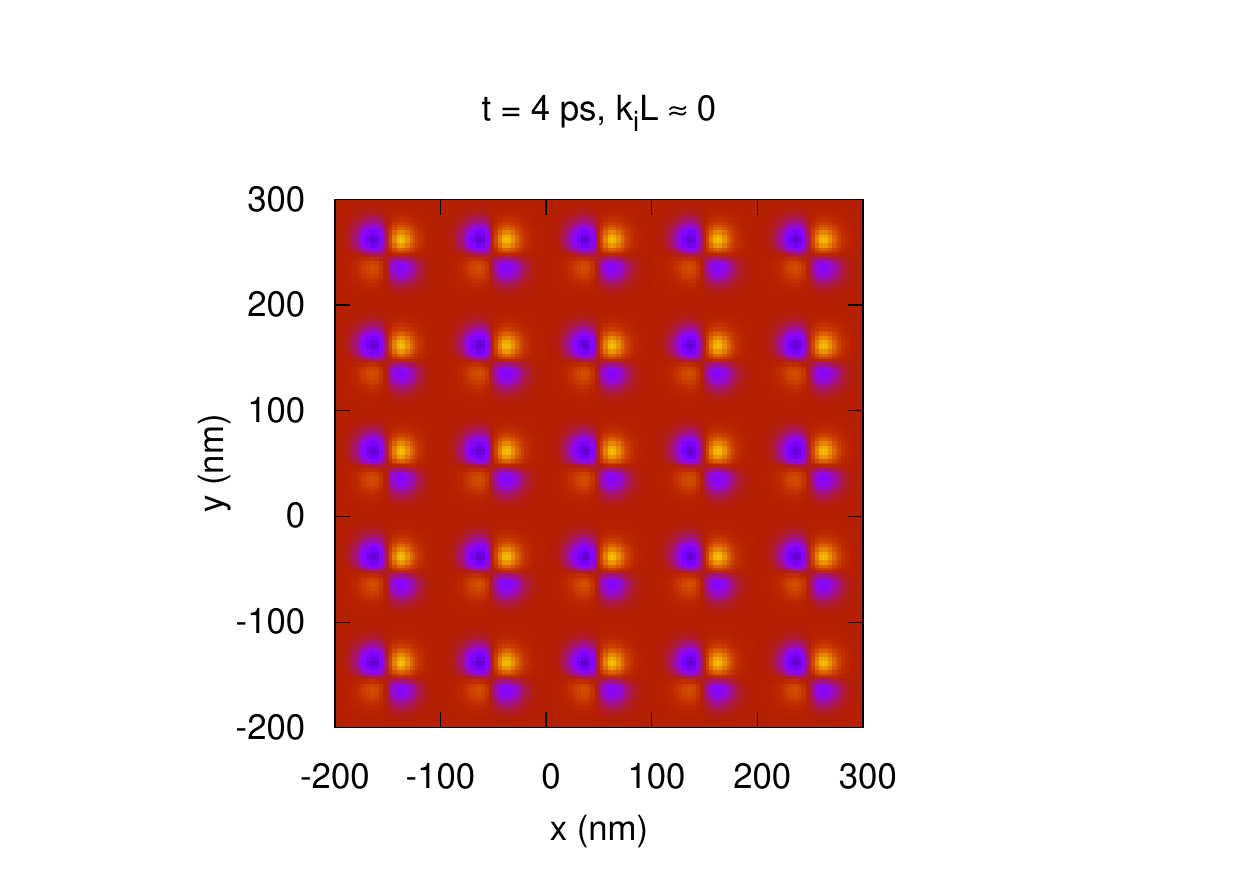}\\
    \vspace*{-1.5cm}
    \includegraphics[width=0.44\textwidth,bb=25 3 300 280]{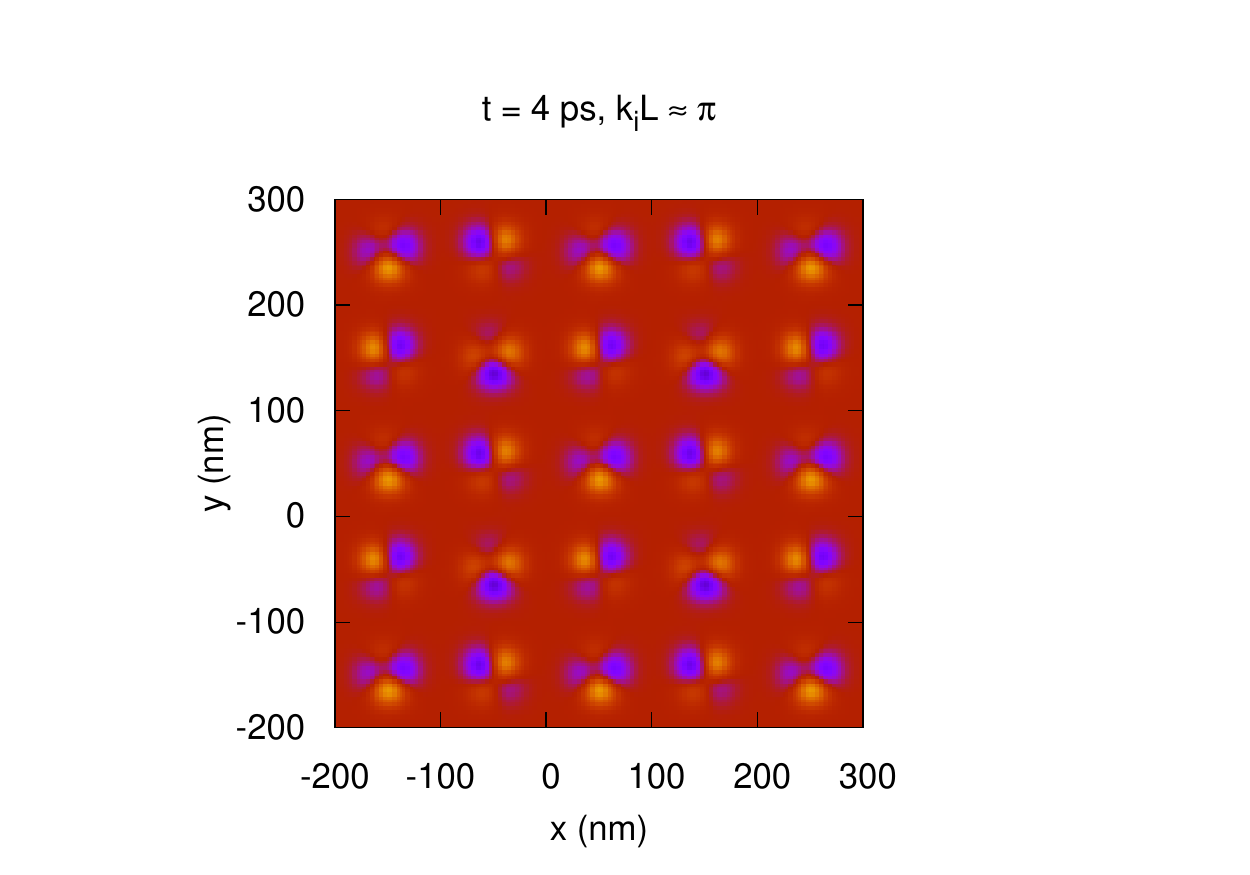}
    \includegraphics[width=0.44\textwidth,bb=25 3 300 280]{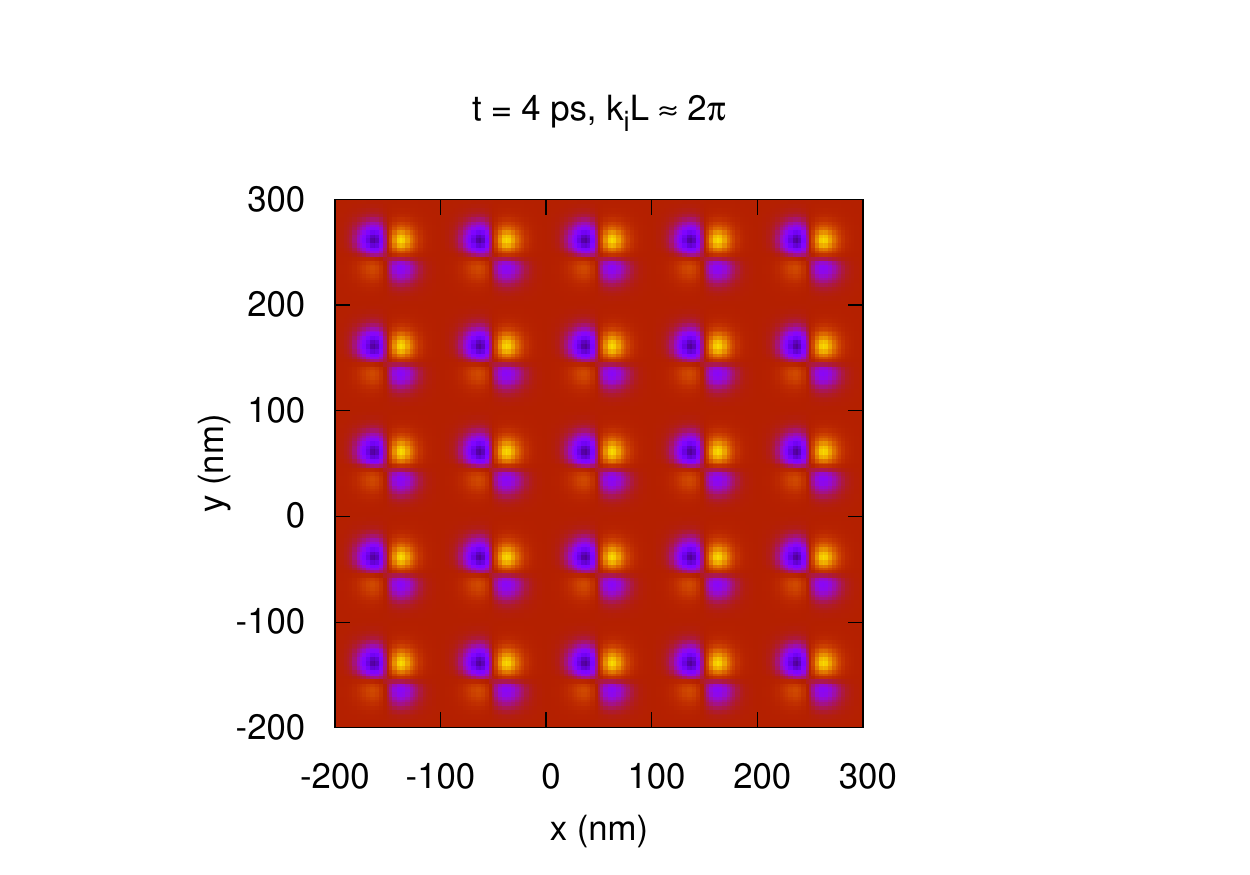}
	\caption{The Fourier power spectra for quadrupole excitation ($Q_2$) for three different
	      	excitation wavevectors (upper left panel), The induced density
	      	$\delta n_e(\bm{r},t) = n_e(\bm{r},t) - n_e(\bm{r},0)$ at $t=4$ ps for
	      	$k_iL\approx 0$ (upper right), $k_iL\approx \pi$ (lower left), and
	      	$k_iL\approx 2\pi$ (lower right) for $i=x,y$.
	      	$pq=4$, $N_e=2$ and $\lambda l = 0$. For the $kL\approx\pi$-case the evolution
	      	of the system is monitored from 0 - 70 ps.
	      	The same range is used for induced density $\delta n_e(\bm{r},t)$ with yellow
	      	color for the upper limit of the color scale and black at its bottom.}
	\label{pq4-Ne2k0pi2pi}
\end{figure*}

In order to employ reasonable cut-offs for the inverse- and the direct lattice sums in the
numerical calculations and investigate collective modes in experimentally relevant regimes
in FIR spectroscopy we generally use $k_iL\approx 0$ with $i=x,y$, but in Fig.\ \ref{pq4-Ne2k0pi2pi} we
compare the excitation of quadrupole modes through $Q_2$ and the induced electron density
for $kL\approx 0,\; \pi$, and $2\pi$ for $pq=4$ and $N_e=2$. The high $kL$ values might be
attainable by some extreme Raman scattering scheme, but we present these results to
show the consistency in the handling of the excitation impulse or wavevector.

In Fig.\ \ref{pq4-Ne2k0pi2pi} the induced density, $\delta n_e(\bm{r},t) = n_e(\bm{r},t) - n_e(\bm{r},0)$,
displays a significant quadrupole modulation. For $N_e = 2$ we see this for $pq=4$ and 3, but not
to the same extent for the lower magnetic flux values $pq = 1$ and 2 (not shown here).
For $kL\approx 0$ we use $k_xL = k_yL = 4\times 10^{-4}$ in the cosine terms in Eq.\ (\ref{phi-ext})
for the excitation pulse, but $(\Delta\bm{k})L = (\bm{\theta} - \bm{\theta}') = 0$ in matrix elements
between different points in the unit inverse lattice (see Eqs.\ (19-20) in \cite{PhysRevB.105.155302}).
Similar choice is used for $kL\approx\pi$ and $2\pi$, i.e.\ $(\bm{\theta} - \bm{\theta}') = \pi$
or $2\pi$, but $kL = \pi + 4\times 10^{-4}$, and $kL = 2\pi + 4\times 10^{-4}$, respectively.
(The small shift, $4\times 10^{-4}$, away from $\pi$ and $2\pi$ is used to avoid $\mu = \pi$ and $\nu = \pi$
in the Ferrari basis \cite{Ferrari90:4598}).

As expected the $Q_2$ excitation peaks for $kL\approx 0$ and $2\pi$ coincide within
numerical accuracy, but the peak for $kL\approx \pi$ is shifted reflecting the completely
different character of that excitation. A look at the dipole excitation $Q_1$ (not shown here)
shows a set of two exactly overlapping peaks, but of different heights as the wavevectors
in the excitation pulse (\ref{phi-ext}) differ.
An excitation pulse (\ref{phi-ext}) with circular polarization leads to relatively
strong contribution of quadrupole and rotational modes, but a pulse with linear
polarization promotes stronger dipole modes in the excitation spectrum.

\begin{figure*}[htb]
	\includegraphics[width=0.48\textwidth,bb=0 50 400 300]{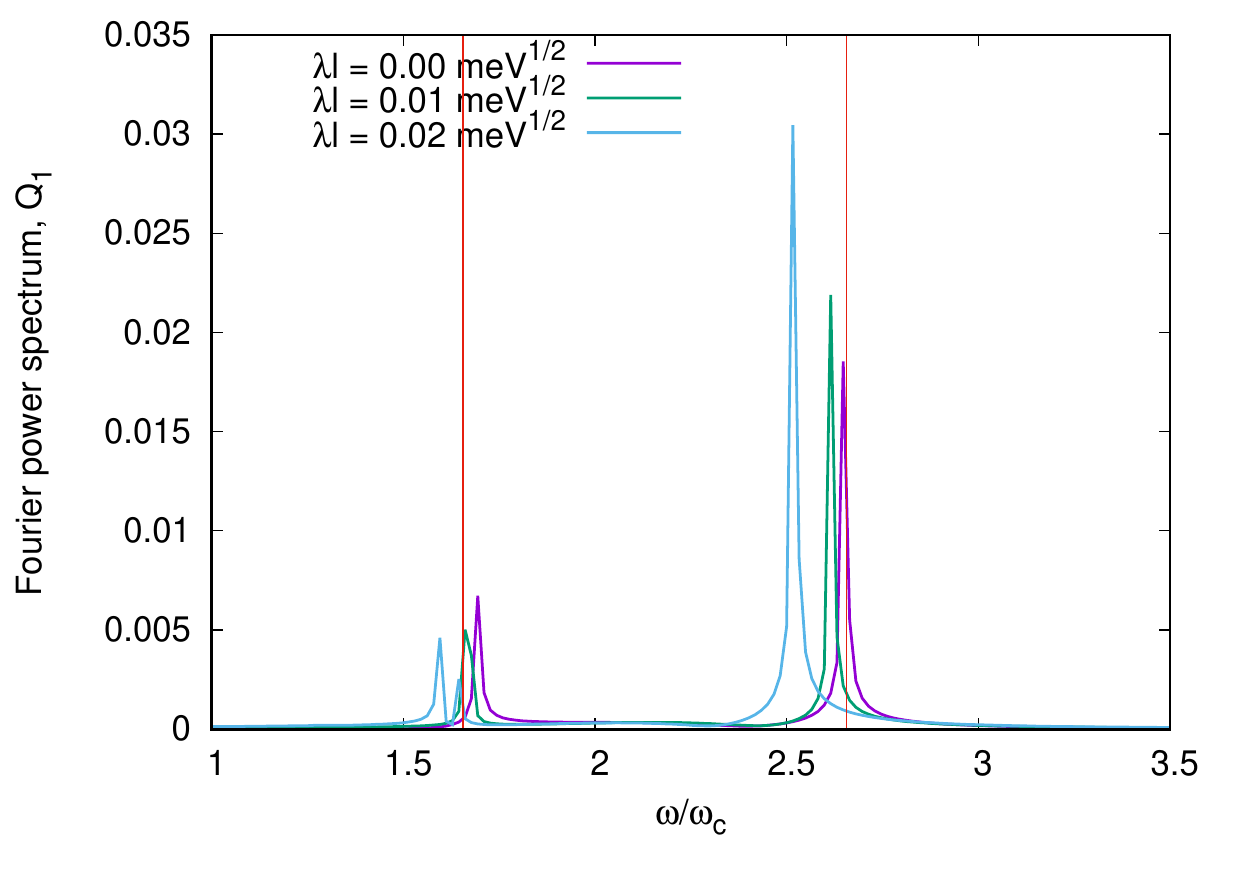}
	\includegraphics[width=0.48\textwidth,bb=0 50 400 300]{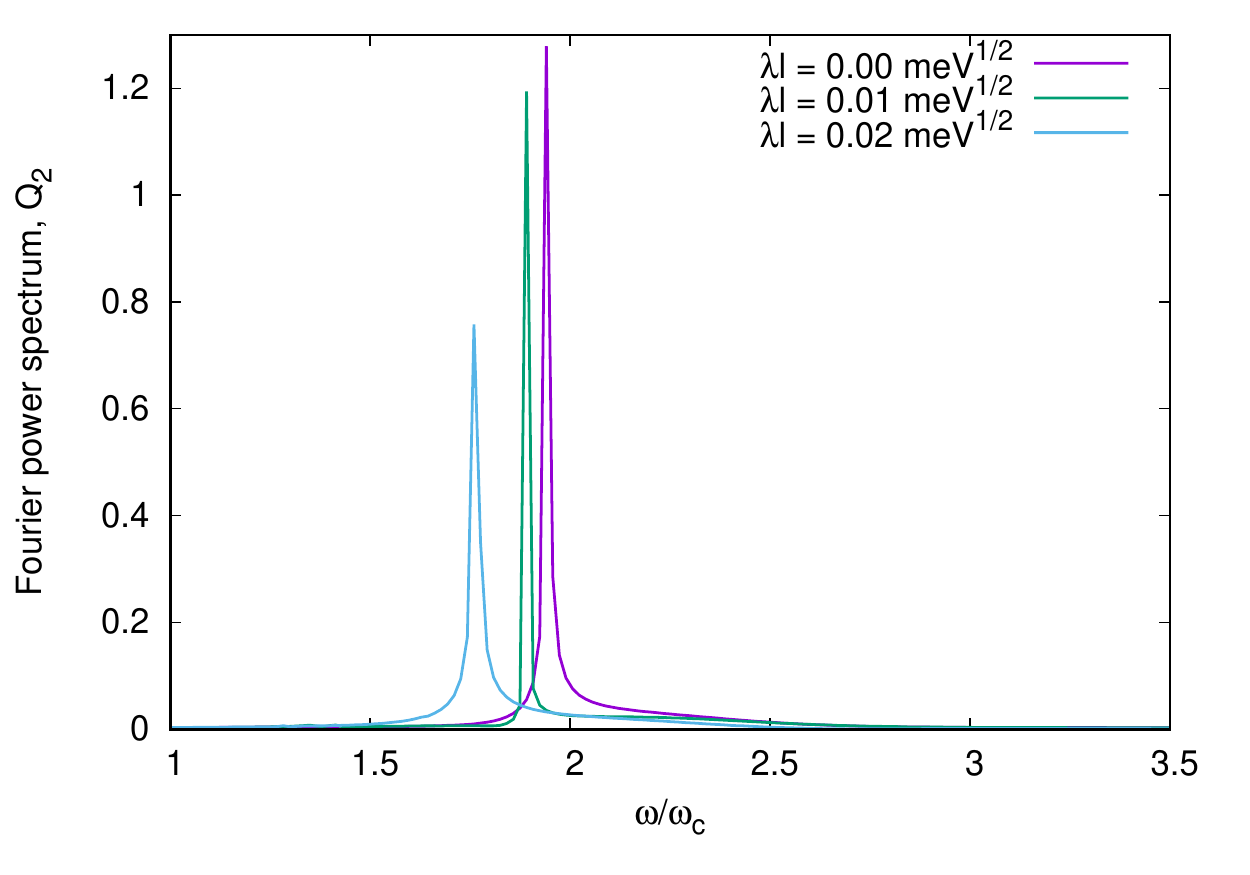}\\
	\vspace*{-0.3cm}
	\includegraphics[width=0.48\textwidth,bb=0 00 400 300]{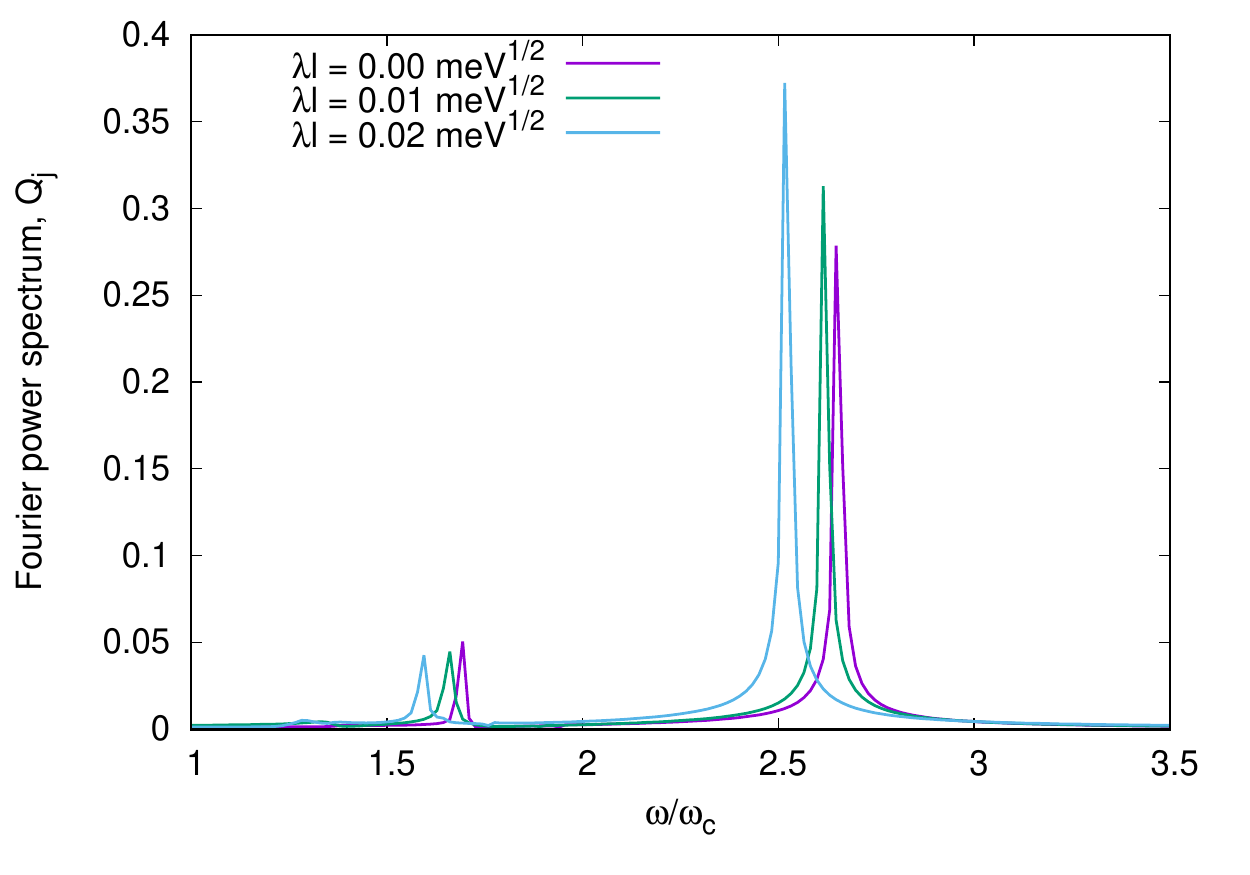}
	\includegraphics[width=0.48\textwidth,bb=0 00 400 300]{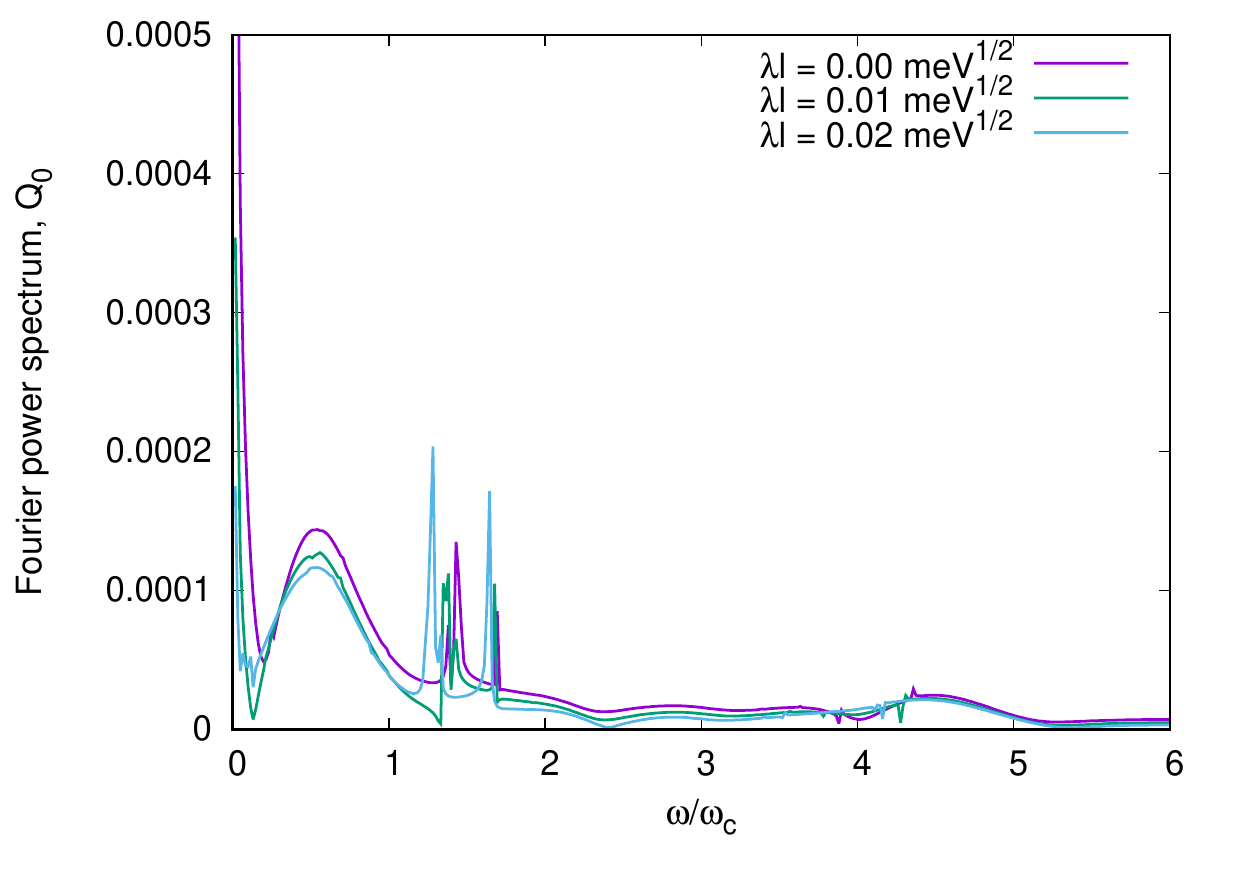}
	\caption{The Fourier power spectra for dipole ($Q_1$) (upper left), quadrupole  ($Q_2$)
		    (upper right), current ($Q_j$) (lower left), and monopole ($Q_0$) (lower right)
		    excitation. $pq=4$, $N_e=2$, $k_xL\approx 0$, $k_yL\approx 0$, and $\hbar\omega_\gamma = 1.0$ meV. The two vertical red lines in the left upper panel indicate the location
		    of the center of mass dipole excitations according to the generalized Kohn theorem
		    for $\lambda l = 0$.}
	\label{pq4-Ne2}
\end{figure*}

Next we investigate the influence of the cavity photon interaction on the excitation spectrum
of the square array of quantum dots. In Fig.\ \ref{pq4-Ne2} we show the results for
$pq = 4$, $N_e = 2$, and $kL\approx 0$ for three different values of the electron-photon
coupling. For two electrons in a dot at this magnetic field we expect the
extended version of the Kohn theorem \cite{Kohn61:1242}
to be almost fulfilled, i.e.\ the electrons ``see'' almost parabolic confinement and the
dipole excitation is a simple oscillation of their center of mass.

In the left upper panel of Fig.\ \ref{pq4-Ne2} the expected location of the dipole excitation
peaks is indicated by two vertical red lines (see \cite{PhysRevB.105.155302} for an
analysis of the periodic confinement potential). Indeed the dipole excitation peaks
coincide very well with the expected Kohn peaks for $\lambda l = 0$, and they are
red-shifted for nonzero electron-photon coupling, like calculations using exact numerical
diagonalization for one photon mode interacting with two electrons in a different
confinement potential have shown \cite{ANDP:ANDP201500298}.
Similar red-shift is found for the quadrupole, $Q_2$, excitation shown in the upper
right panel of Fig.\ \ref{pq4-Ne2} as a function of the electron-photon interaction
strength $\lambda l$. The location of the quadrupole peaks in relation with the
dipole peaks is in accordance with earlier results for quantum dots
\cite{PhysRevLett.54.1710,Gudmundsson91:12098,PhysRevB.105.155302}.
The location of the rotational excitation peaks, $Q_j$, corresponds with results from
a real-time Hartree approximation \cite{PhysRevB.105.155302}, i.e.\ they coincide with
the dipole peaks and their red-shifts are in accordance with the red-shift of the dipole
peaks. In accordance with the fact that the dipole oscillations are mainly center of mass
(CM) oscillations the monopole excitation spectrum seen in the right lower panel of
Fig.\ \ref{pq4-Ne2} only shows very low peaks and broad features.

The concurrent increase in the oscillator strength of the upper dipole peak
as it is red-shifted with increasing $\lambda l$ can be understood as an increasing
weight of a bulk mode as the polarization of the charge density is increased, and the
opposite effect is seen for the lower dipole peak, the ``edge mode''.
With this in mind we can understand the decrease of the strength of the quadrupole mode
as an effect of increased screening of the angular dependent structure
facilitated by the charge polarizability caused by the electron-photon interaction.

\begin{figure*}[htb]
	\includegraphics[width=0.48\textwidth,bb=0 50 400 300]{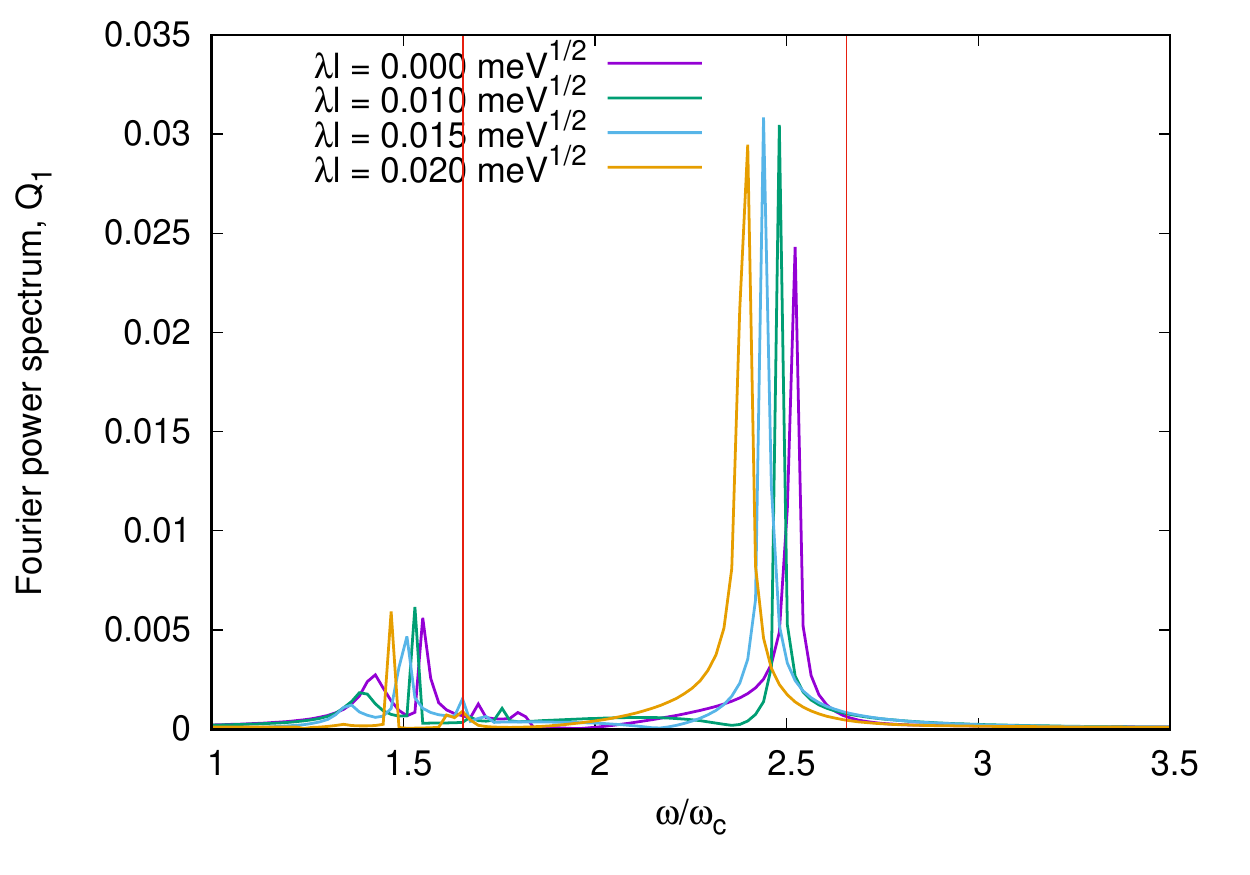}
	\includegraphics[width=0.48\textwidth,bb=0 50 400 300]{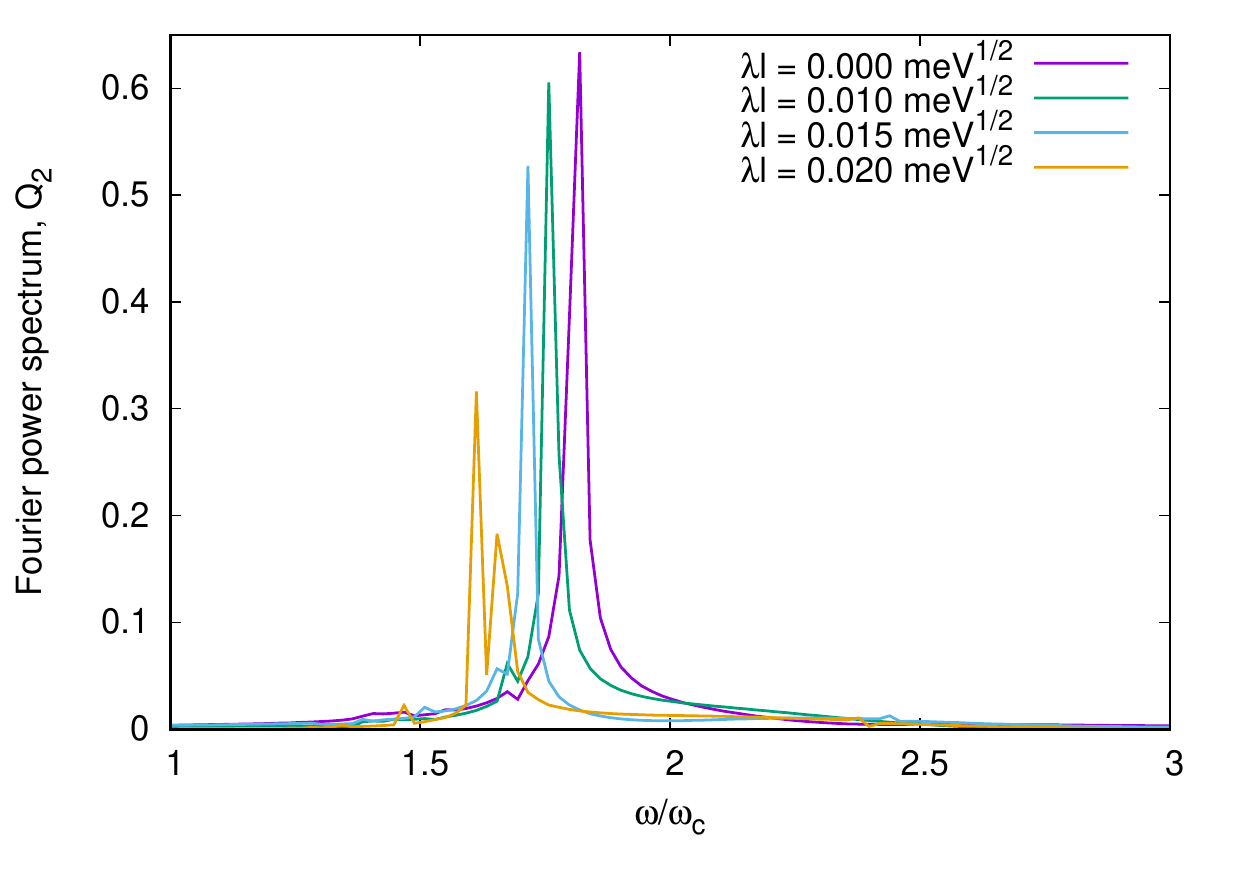}\\
	\vspace*{-0.3cm}
	\includegraphics[width=0.48\textwidth,bb=0 00 400 300]{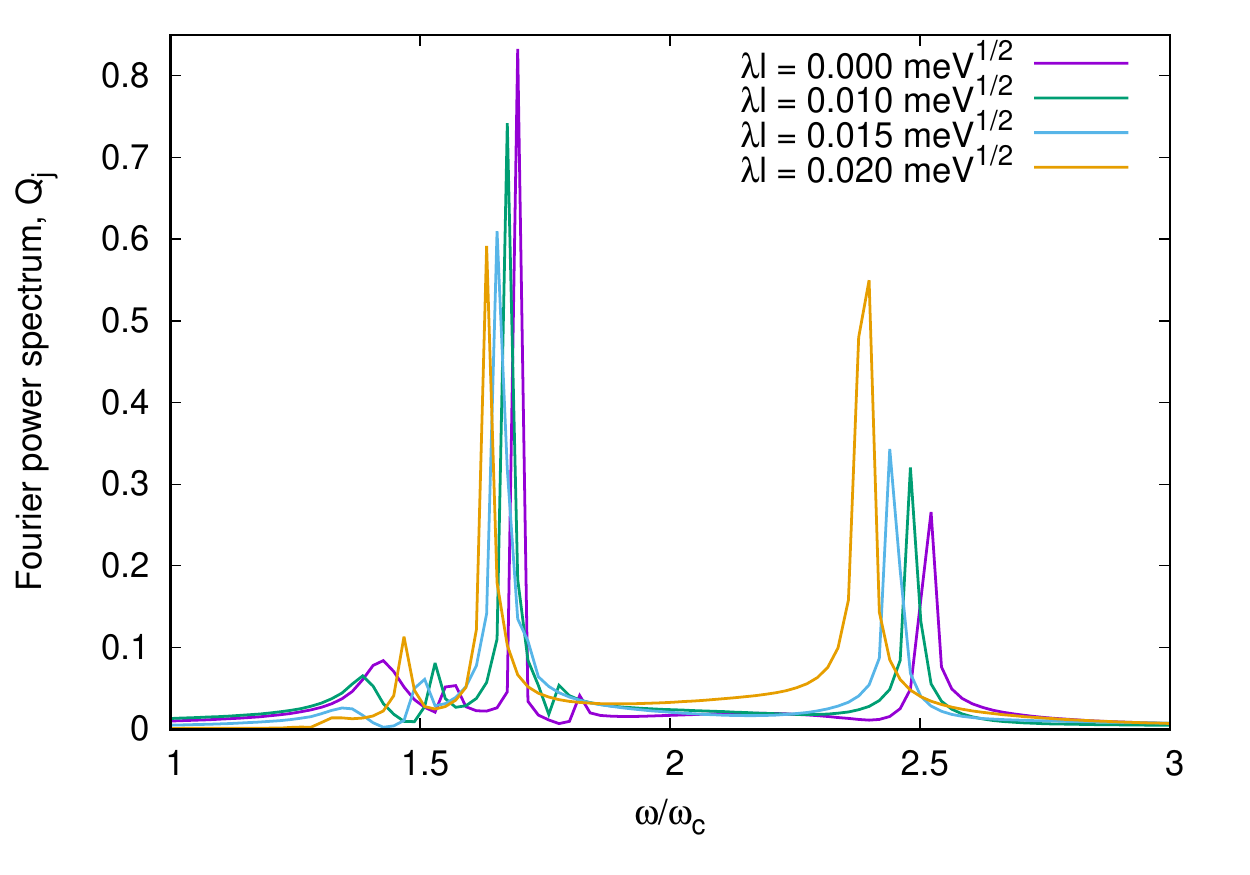}
	\includegraphics[width=0.48\textwidth,bb=0 00 400 300]{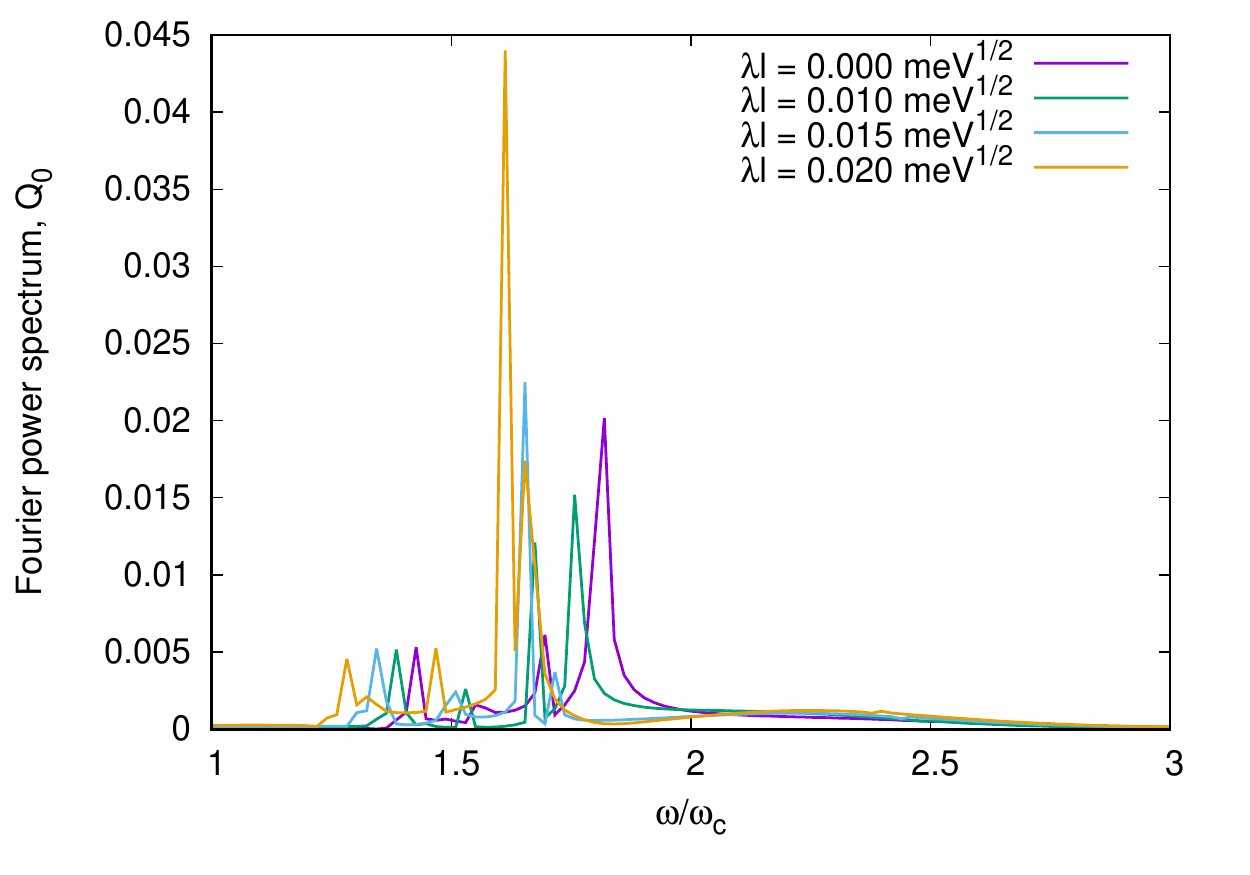}
	\caption{The Fourier power spectra for dipole ($Q_1$) (upper left), quadrupole  ($Q_2$)
		(upper right), current ($Q_j$) (lower left), and monopole ($Q_0$) (lower right)
		excitation. $pq=4$, $N_e=2$, $k_xL\approx \pi$, $k_yL\approx \pi$, and $\hbar\omega_\gamma = 1.0$ meV. The two vertical red lines in the left upper panel indicate the location
		of the center of mass dipole excitations according to the generalized Kohn theorem
		for $\lambda l = 0$.}
	\label{pq4-Ne2-pi}
\end{figure*}

In Fig.\ \ref{pq4-Ne2-pi} we present the excitation spectra for an array of quantum
dots with the same number of electrons per unit cell and magnetic flux ($pq = 4$) as was
seen in Fig.\ \ref{pq4-Ne2}, but now the excitation has a much larger wavevector,
$kL\approx\pi$ which breaks the condition for a simple harmonic excitation described
by the Kohn theorem.

Still, the electron-photon interaction red-shifts all the excitation peaks like before,
but now extra features can be seen for the largest coupling, and clearly stronger
monopole excitations can be seen reflecting some kinds of breathing modes.
This last point is related to the excitation of monopole modes in
Raman scattering \cite{Schueller96:xx,Steinebach99:10240,Steinebach00:15600}.

\begin{figure*}[htb]
	\includegraphics[width=0.48\textwidth,bb=0 50 400 300]{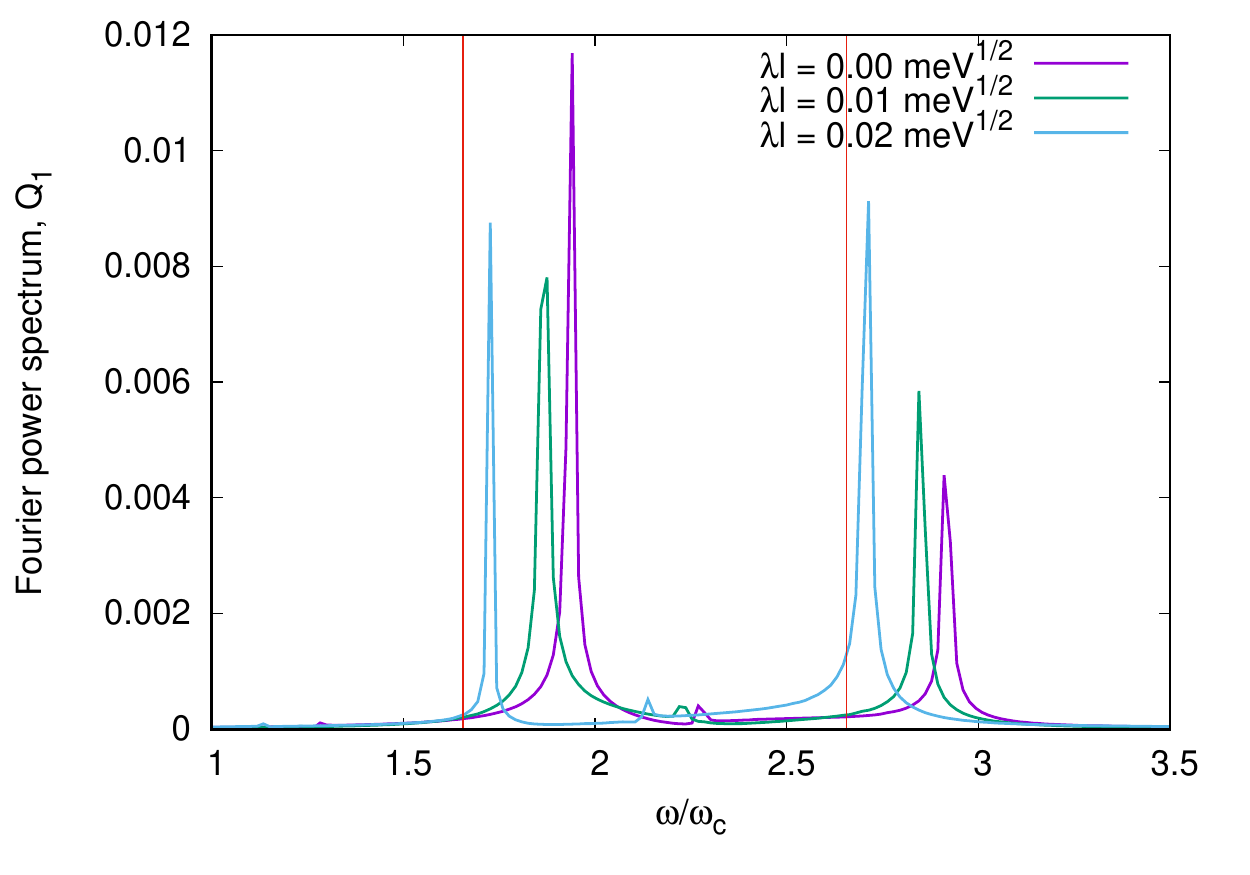}
	\includegraphics[width=0.48\textwidth,bb=0 50 400 300]{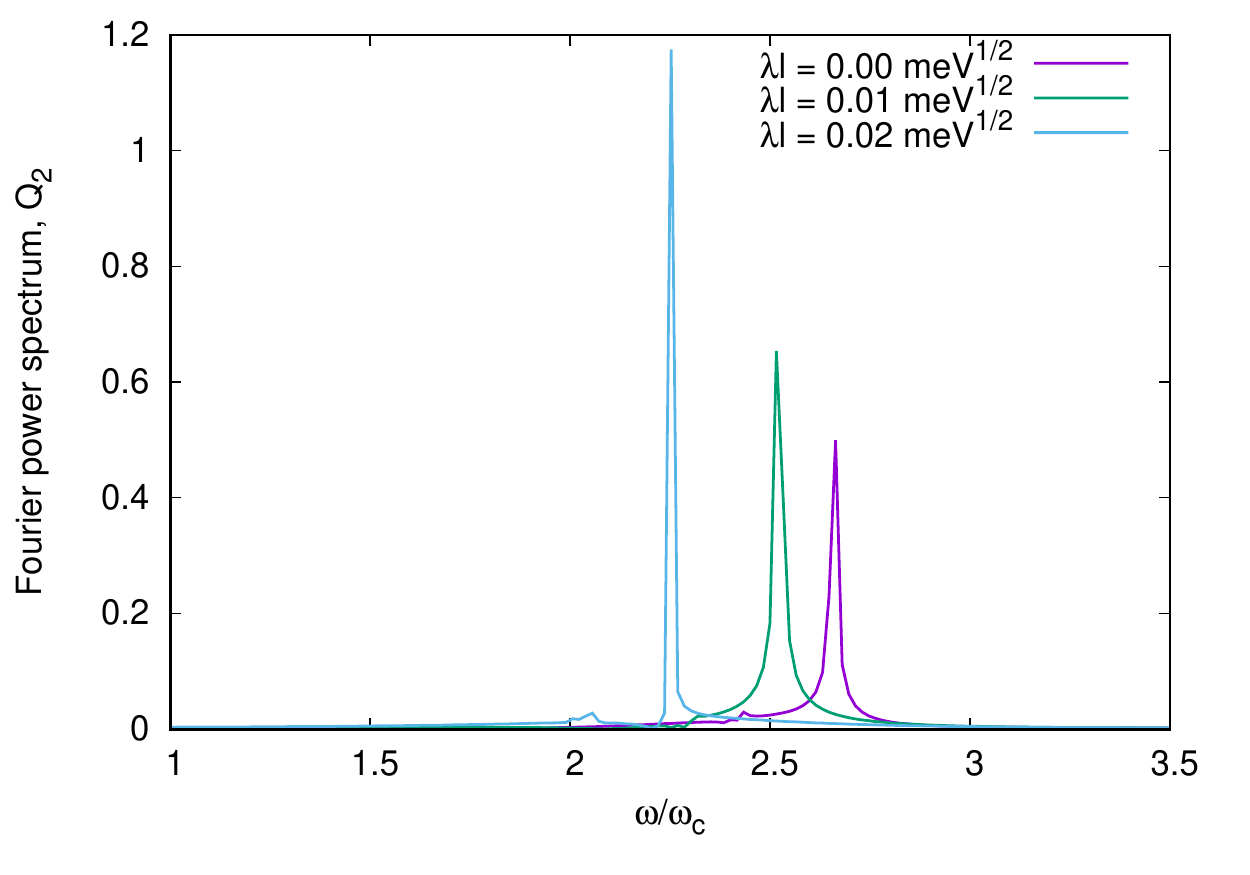}\\
	\vspace*{-0.3cm}
	\includegraphics[width=0.48\textwidth,bb=0 00 400 300]{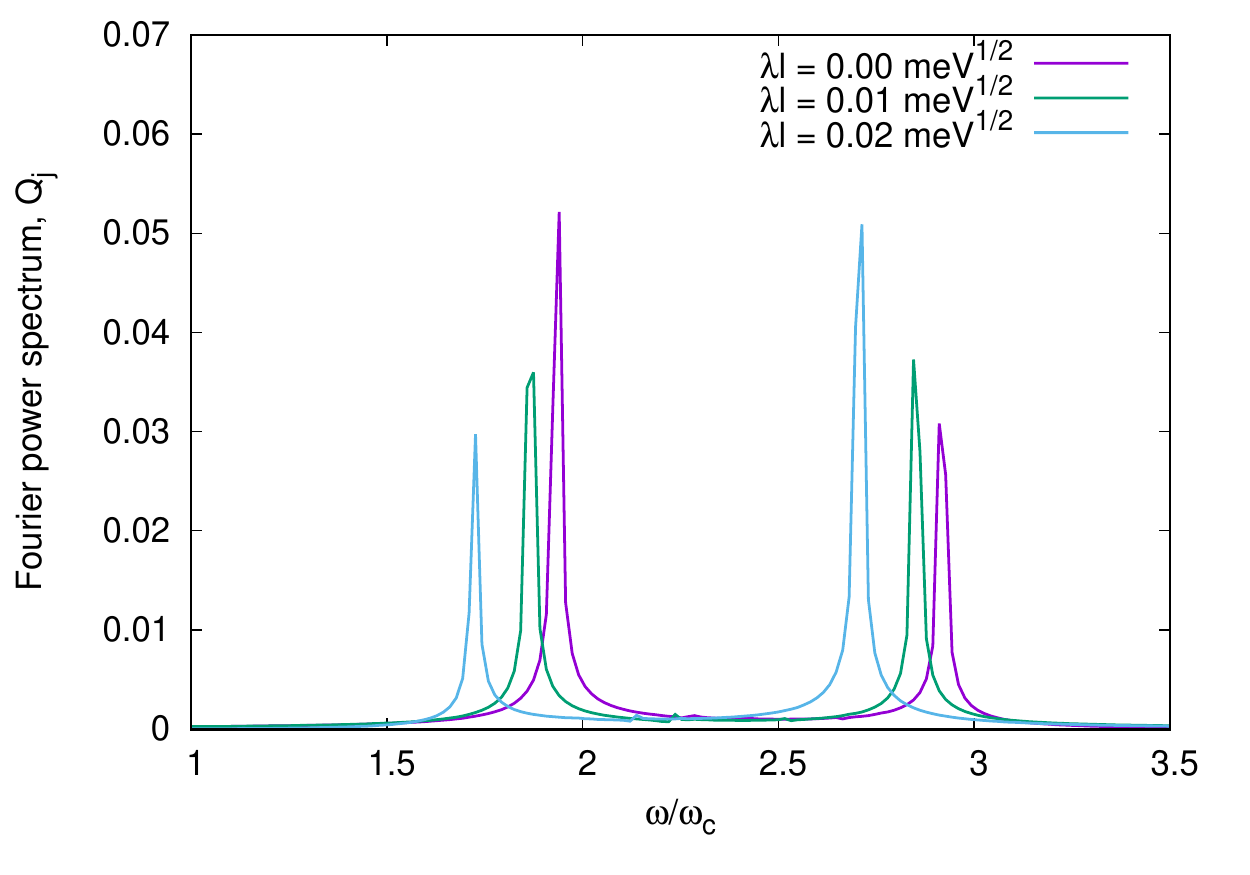}
	\includegraphics[width=0.48\textwidth,bb=0 00 400 300]{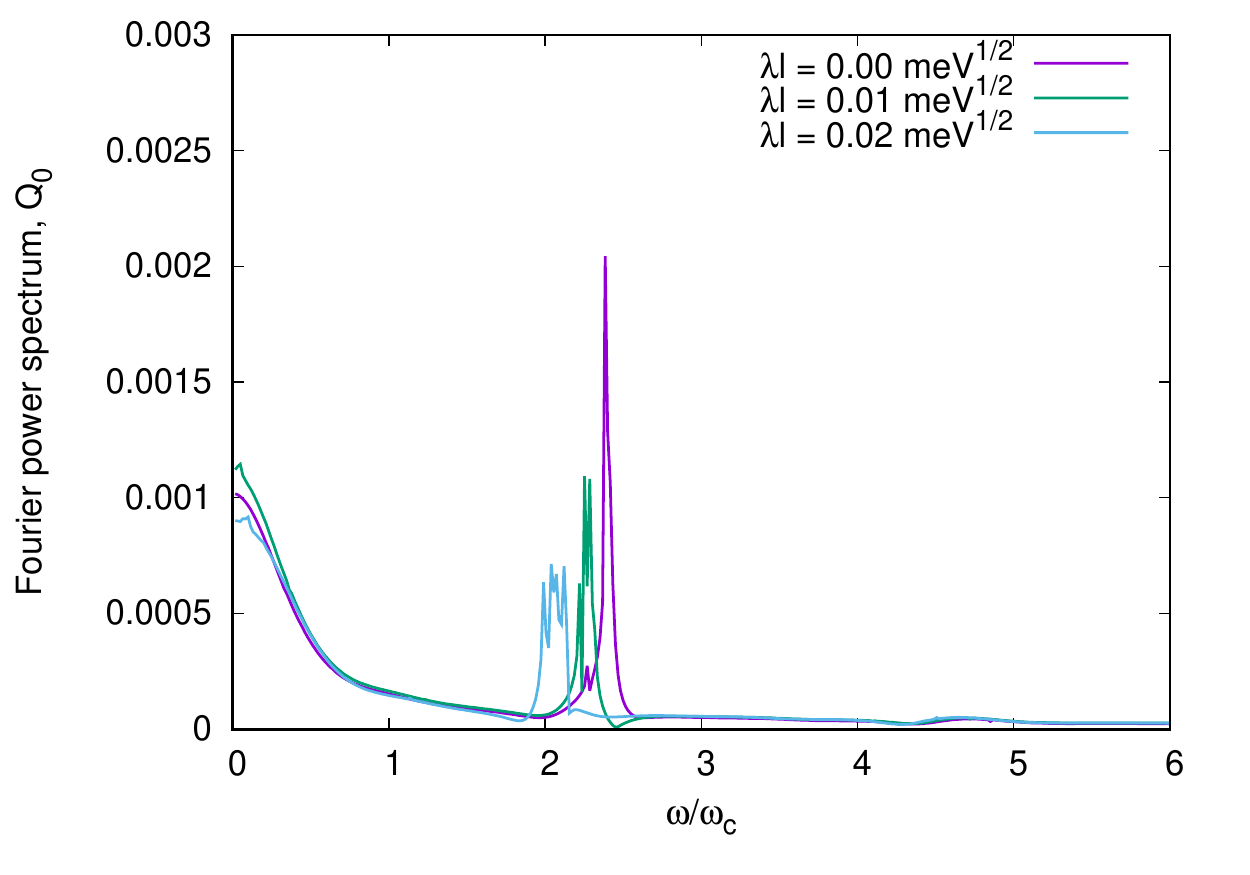}
	\caption{The Fourier power spectra for dipole ($Q_1$) (upper left), quadrupole  ($Q_2$)
	    	(upper right), current ($Q_j$) (lower left), and monopole ($Q_0$) (lower right)
	     	excitation. $pq=4$, $N_e=1$, $k_xL\approx 0$, $k_yL\approx 0$, and $\hbar\omega_\gamma = 1.0$ meV. Note the clear blue shift of the dipole excitation and the
	     	higher amplitudes of the quadrupole peak as the strength of
	     	the electron-photon $\lambda l$ increases. The two vertical red lines in the left
	     	upper panel indicate the location of the center of mass dipole excitations
	     	according to the generalized Kohn theorem for $\lambda l = 0$.}
	\label{pq4-Ne1}
\end{figure*}

So, what happens as we go back to the parameters used for the results presented in
Fig.\ \ref{pq4-Ne2}, but now instead of 2 electrons in a dot we set $N_e = 1$?
The first thought might be that as the electron is deeper in the dot potential
we must have even a better adherence to the Kohn theorem. The results presented
in Fig.\ \ref{pq4-Ne1} show is that this simple line of thoughts is not adequate
to describe the situation.
We are forgetting an important fact, the array structure with rather short period and the
role of the exchange force that is maximized by the electron number being an odd low
integer. The generalized Kohn theorem states that only CM-oscillations are excited
by an external electric field with wavelength much larger than the dot if the electrons
are parabolically confined in it, independent of the interaction between the electrons
as long as they are pair interactions, only depending on their relative positions.
For two electrons in a dot the intra-dot Coulomb interaction is strong and creates
no problem regarding the Kohn theorem, but the inter-dot interaction makes a small
deviation that is seen in Fig.\ \ref{pq4-Ne1}, especially regarding the location of
the lower $Q_1$ excitation peak. For $N_e = 1$ the situation is different, the exchange
force creates an enhanced gap between the occupied lowest subband and the unoccupied higher subbands
and all peaks in the excitation spectra are blue-shifted by the strong Coulomb exchange interaction.
On top of this the electron-photon interaction adds a red-shift like earlier.

\begin{figure*}[htb]
	\includegraphics[width=0.48\textwidth,bb=0 50 400 300]{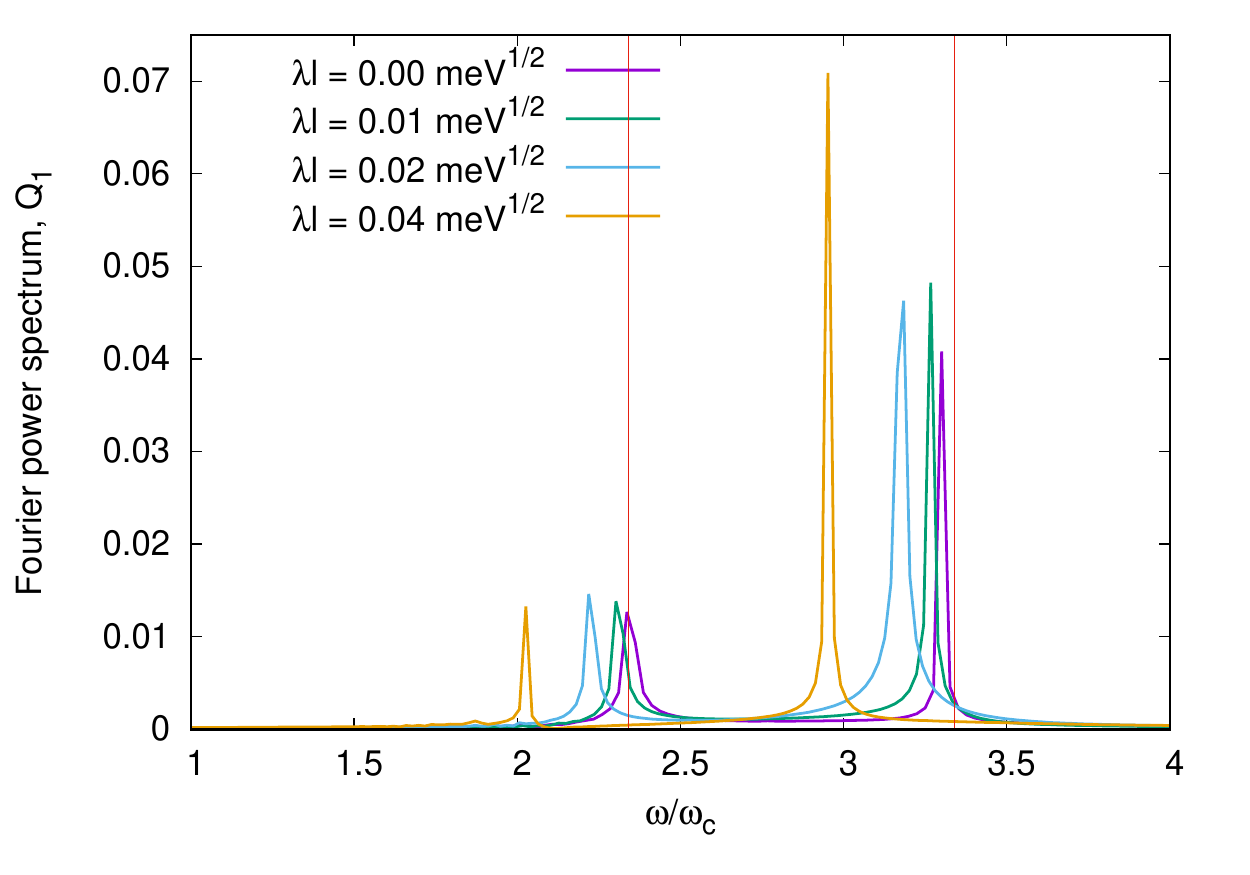}
	\includegraphics[width=0.48\textwidth,bb=0 50 400 300]{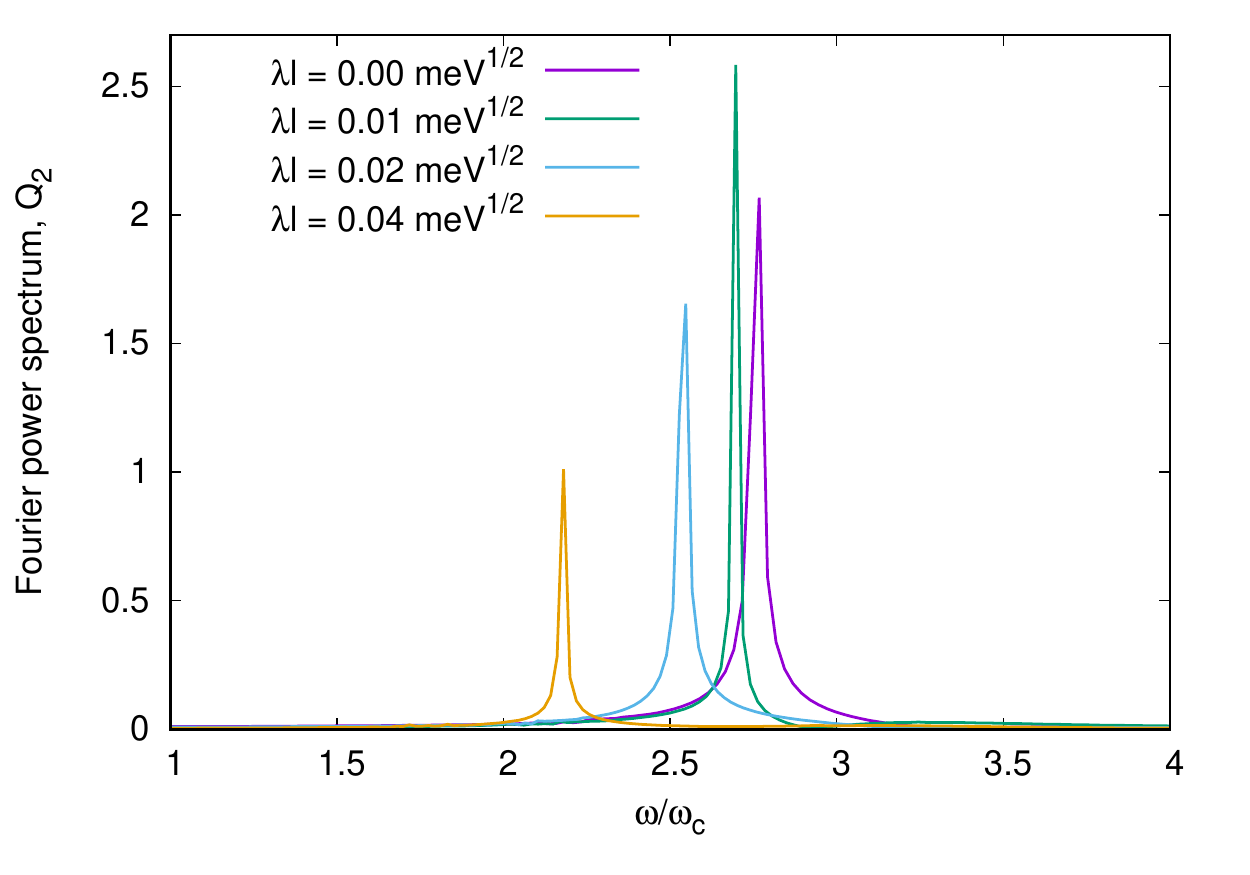}\\
	\vspace*{-0.3cm}
	\includegraphics[width=0.48\textwidth,bb=0 00 400 300]{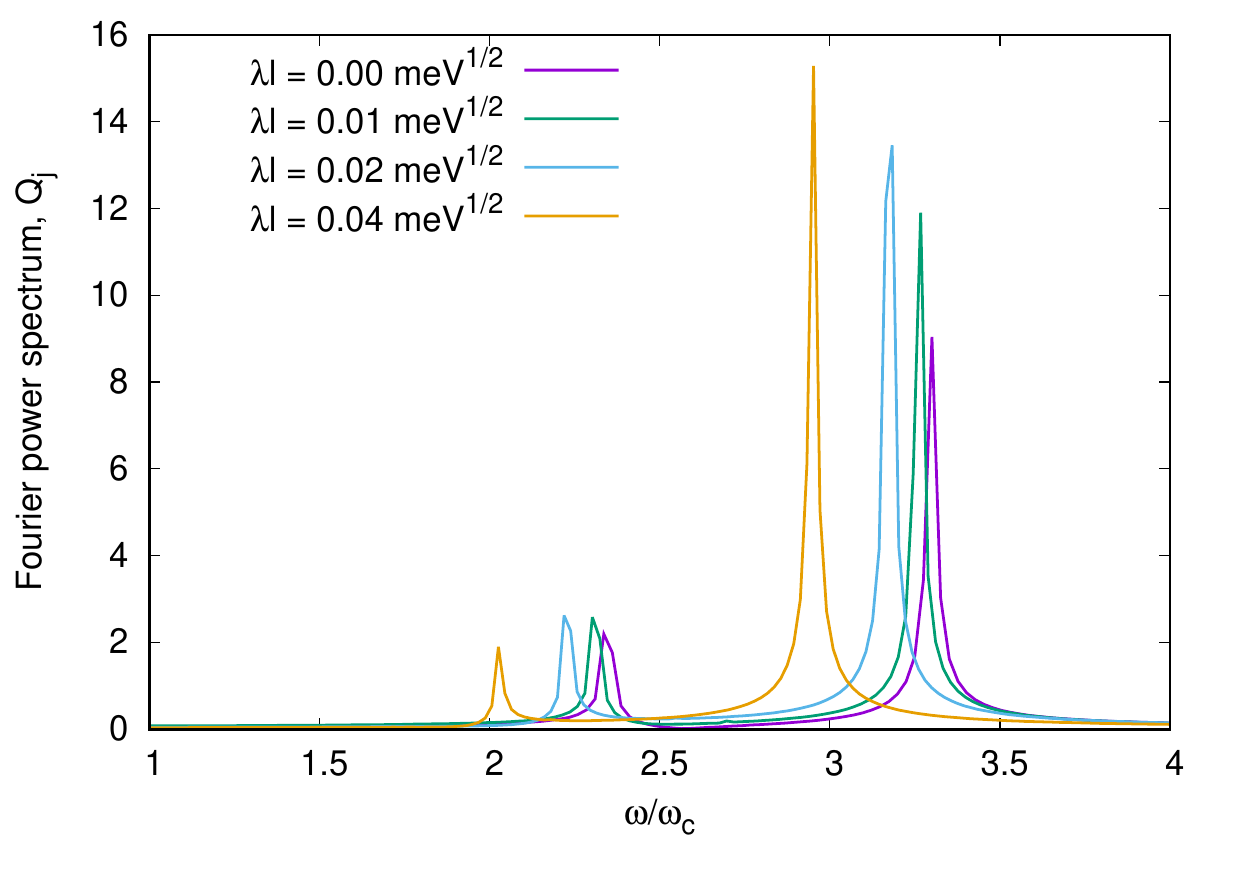}
	\includegraphics[width=0.48\textwidth,bb=0 00 400 300]{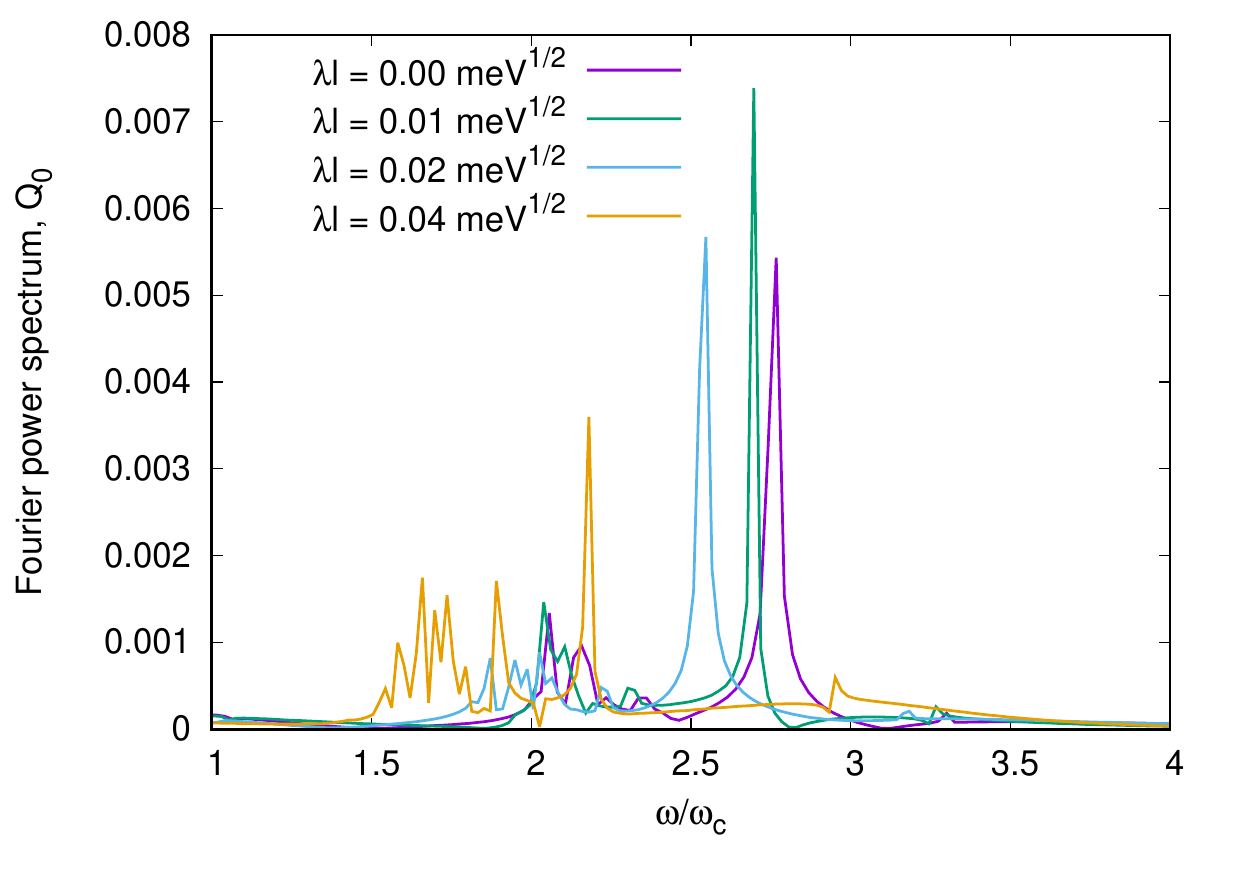}
	\caption{The Fourier power spectra for dipole ($Q_1$) (upper left), quadrupole  ($Q_2$)
	    	(upper right), current ($Q_j$) (lower left), and monopole ($Q_0$) (lower right)
	    	excitation. $pq=3$, $N_e=2$, $k_xL\approx 0$, $k_yL\approx 0$, and $\hbar\omega_\gamma = 1.0$ meV. The two vertical red lines in the left	upper panel indicate the location of
	    	the center of mass dipole excitations according to the generalized Kohn theorem
	    	for $\lambda l = 0$.}
	\label{pq3-Ne2}
\end{figure*}

Very similar behavior is found for the excitation spectra for $N_e = 2$ and the lower
magnetic field corresponding to $pq = 3$ displayed in Fig.\ \ref{pq3-Ne2}.
As the magnetic length is now a bit longer and the electrons ``feel'' more of the
nonparabolic part of the confinement potential leading to a slightly more deviation
from the simple Kohn results. Still, the weak monopole excitation spectrum seen in
the lower right panel in Fig.\ \ref{pq3-Ne2} indicates only a slight deviation from
the Kohn results judging from the main peak, but the several low intensity peaks in
the lower energy range hint at a small overlapping of the charge density of the dots.
Again, the electron-photon interaction leads to a red-shift of all
the excitation peaks with similar characteristics as was seen for the $pq = 4$ cases.

\begin{figure*}[htb]
	\includegraphics[width=0.48\textwidth,bb=0 50 400 300]{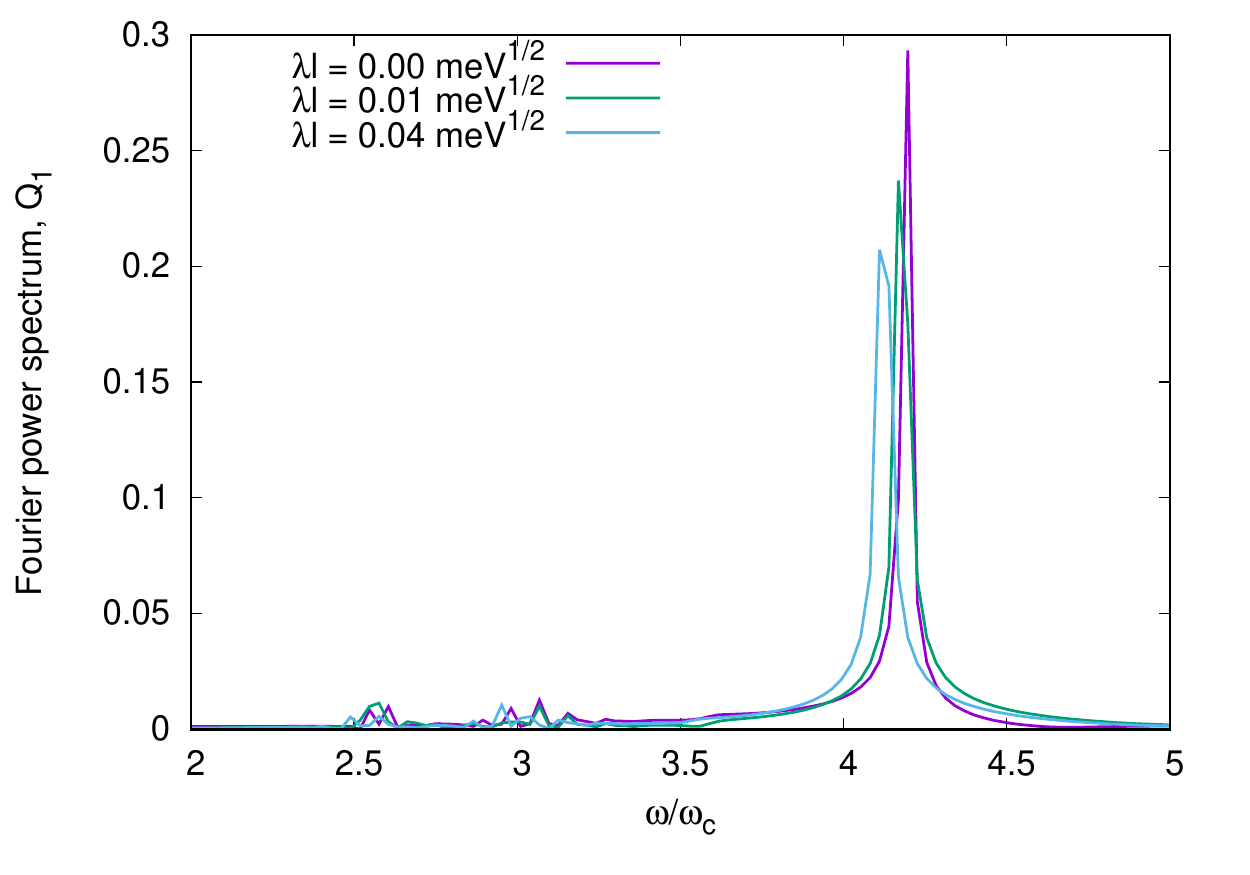}
	\includegraphics[width=0.48\textwidth,bb=0 50 400 300]{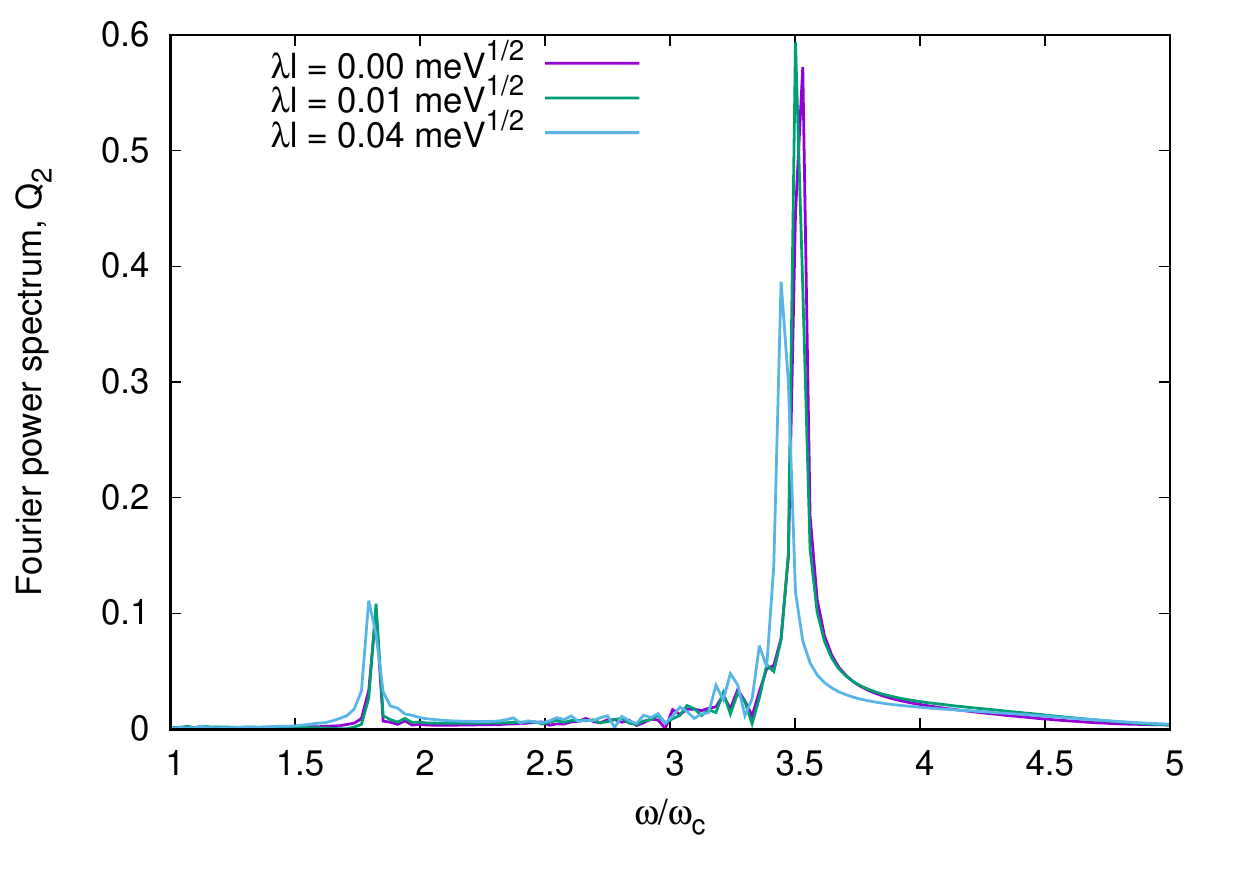}\\
	\vspace*{-0.3cm}
	\includegraphics[width=0.48\textwidth,bb=0 00 400 300]{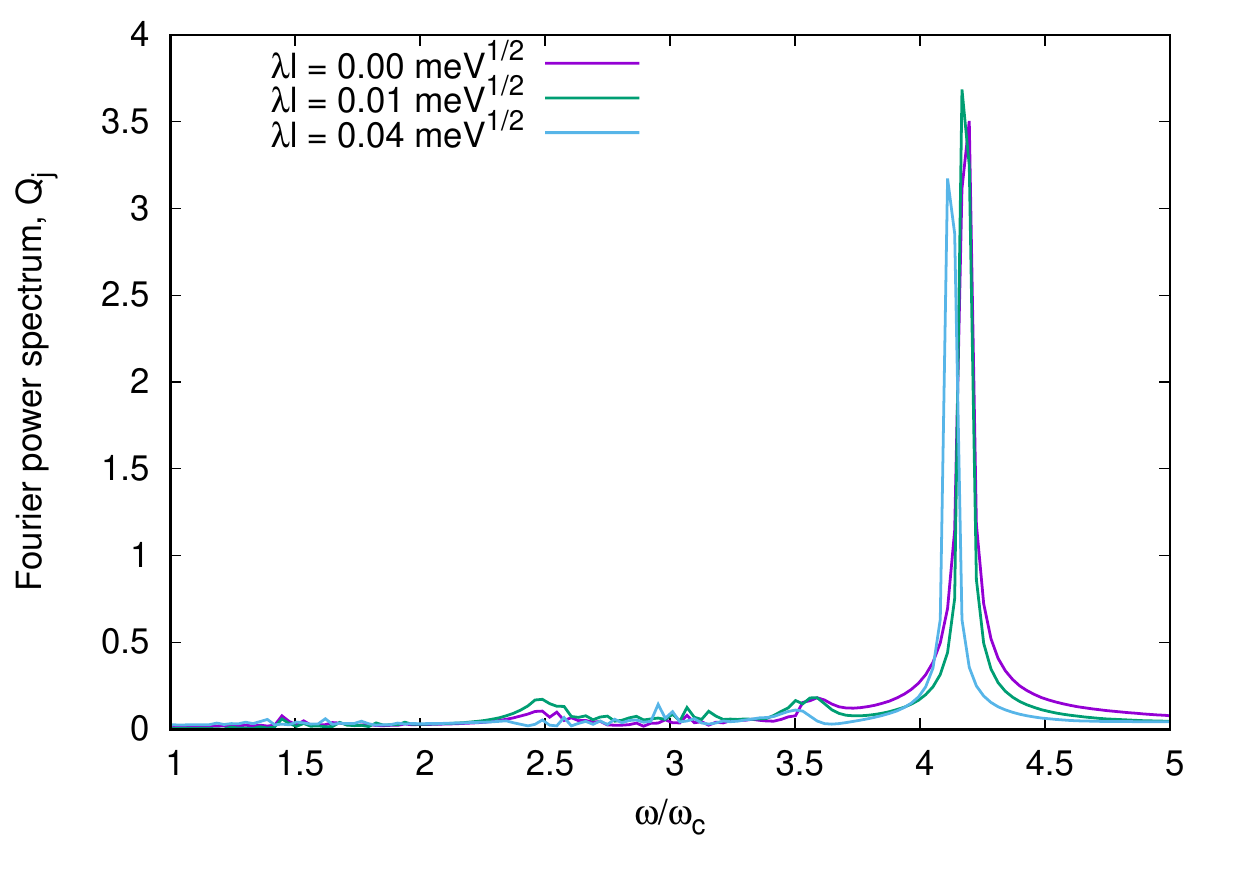}
	\includegraphics[width=0.48\textwidth,bb=0 00 400 300]{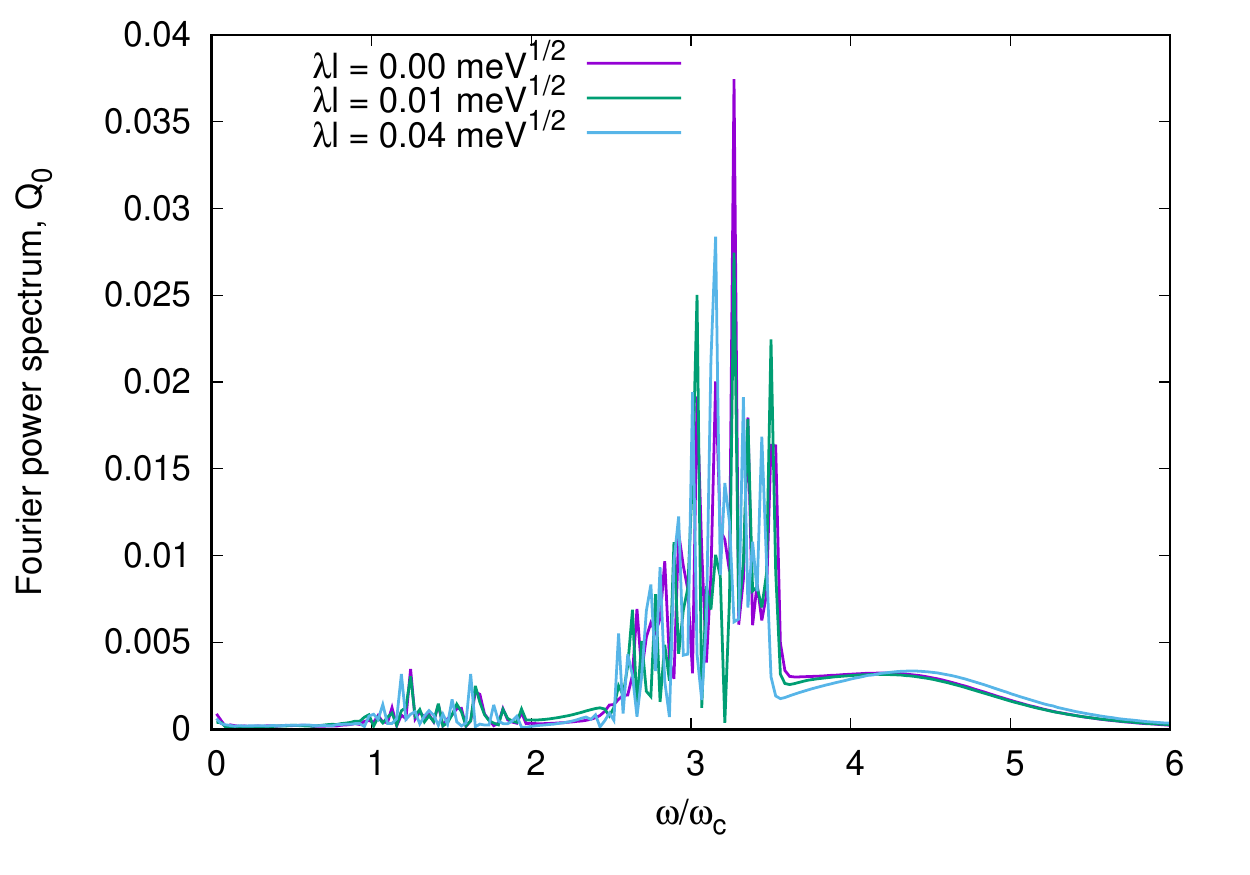}
	\caption{The Fourier power spectra for dipole ($Q_1$) (upper left), quadrupole  ($Q_2$)
		    (upper right), current ($Q_j$) (lower left), and monopole ($Q_0$) (lower right)
	    	excitation. $pq=2$, $N_e=3$, $k_xL\approx 0$, $k_yL\approx 0$, and $\hbar\omega_\gamma = 1.0$ meV.}
	\label{pq2-Ne3}
\end{figure*}

In order to observe the effects of the electron-photon interaction on the excitation
spectrum for the system when it is not protected by the Kohn theorem we choose two cases.
First, we show in Fig.\ \ref{pq2-Ne3}
the excitation spectra for the lower magnetic field corresponding to $pq = 2$ and
$N_e = 3$, where electron charge density is not vanishingly small between the dots in
the array. Accordingly, the dipole spectrum (upper left panel) has one strong peak
and two much lower peak regions caused by the splitting of the lower Kohn peak by the
square symmetry of the array \cite{PhysRevB.60.16591}. The red-shift caused by the
electron-photon interaction is smaller just like the change in the orbital magnetization
has been seen to be smaller for a higher number of electrons \cite{PhysRevB.106.115308}.

\begin{figure*}[htb]
	\includegraphics[width=0.48\textwidth,bb=0 50 400 300]{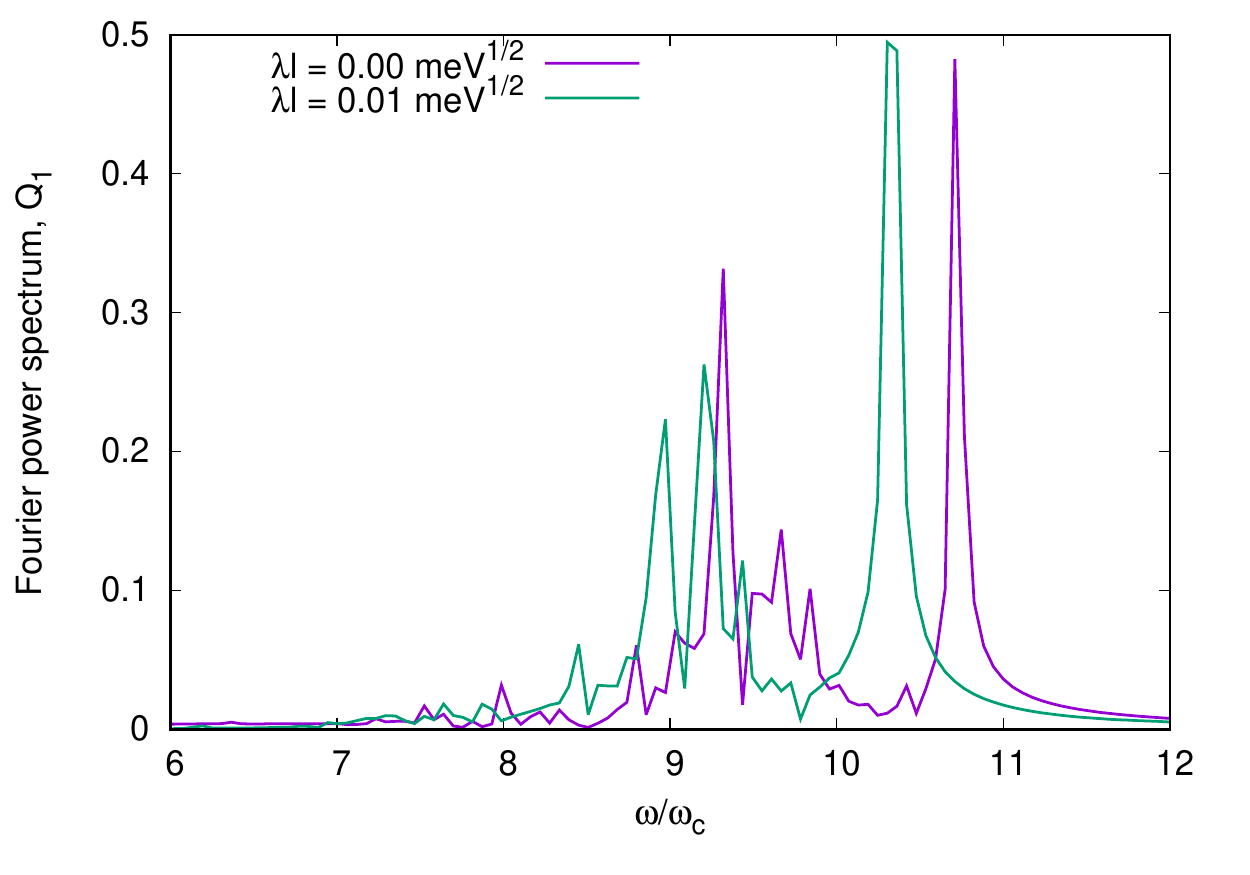}
	\includegraphics[width=0.48\textwidth,bb=0 50 400 300]{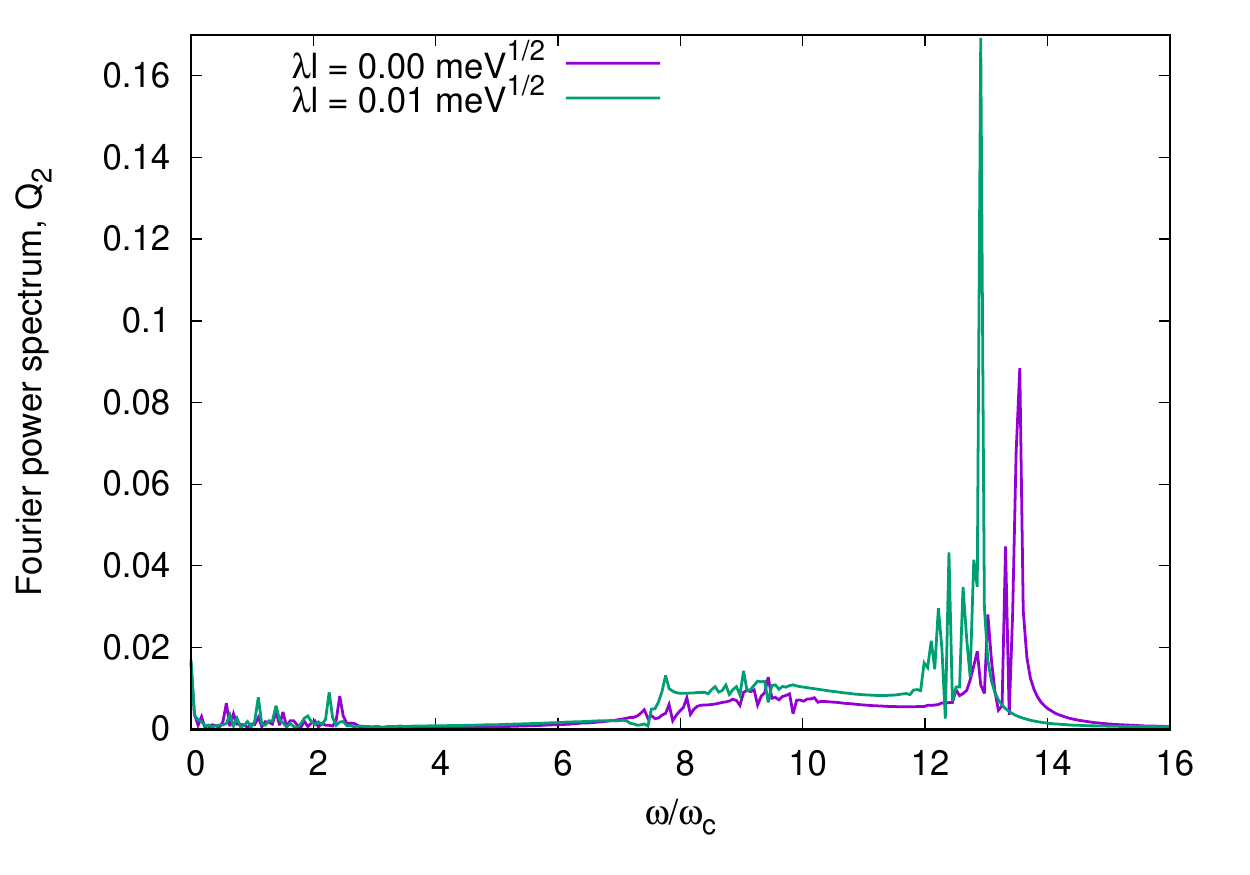}\\
	\vspace*{-0.3cm}
	\includegraphics[width=0.48\textwidth,bb=0 00 400 300]{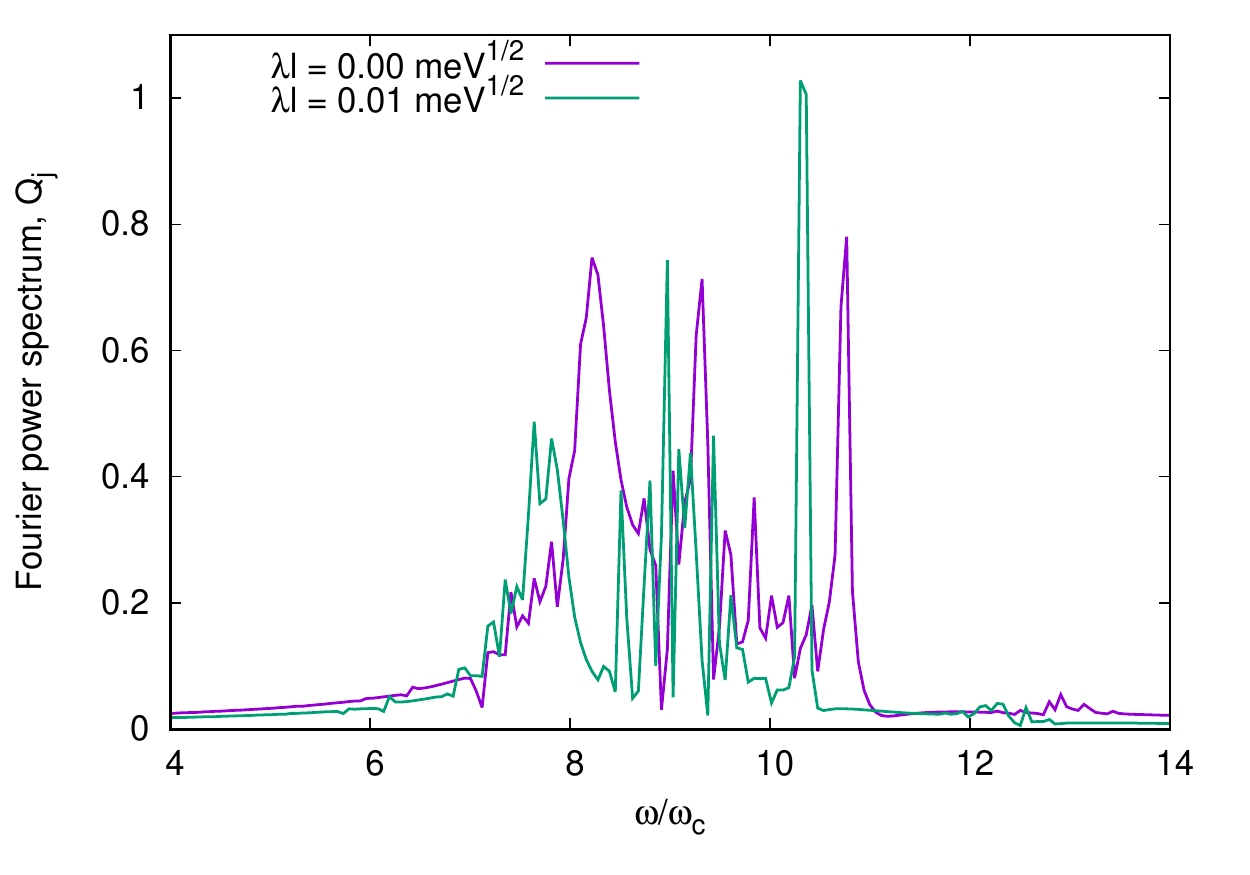}
	\includegraphics[width=0.48\textwidth,bb=0 00 400 300]{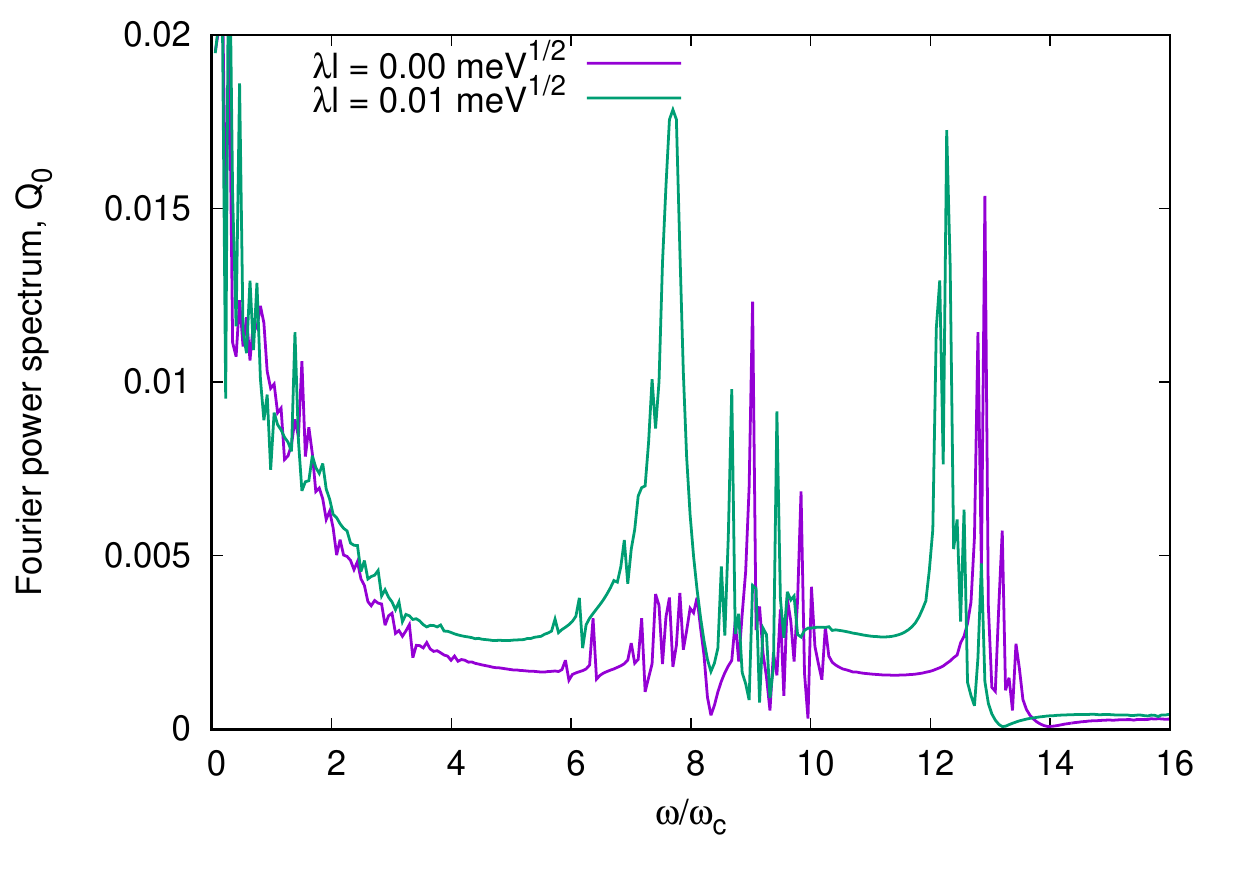}
	\caption{The Fourier power spectra for dipole ($Q_1$) (upper left), quadrupole  ($Q_2$)
		    (upper right), current ($Q_j$) (lower left), and monopole ($Q_0$) (lower right)
		    excitation. $pq=1$, $N_e=1$, $k_xL\approx 0$, $k_yL\approx 0$, and $\hbar\omega_\gamma = 1.0$ meV.}
	\label{pq1-Ne1}
\end{figure*}

The second case, where the Kohn theorem does not apply is presented in Fig.\ \ref{pq1-Ne1},
where $pq = 1$ and $N_e = 1$, so one electron in each dot, or unit cell of the lattice,
at a low magnetic field.
The dipole spectrum (upper left panel) shows a clear upper peak and then a group of
peaks lower in energy. The induced density (not shown here) still shows clear CM oscillations,
but with some modification caused by the nonparabolic confinement potential felt by the electrons.
Importantly, for the low number of electrons, just one in each dot, we see again a large
red-shift caused by a weak electron-photon interaction, but at the same time there is a large
allover blue-shift due to the strong interdot exchange force. The Kohn peaks for the dipole would
be approximately located at 7.8 and 8.8 $\omega /\omega_c$. The complex lower part of the
dipole excitation spectrum reflects both the weakening of the confinement potential away from
a simple parabolic confinement, and at the same time the influence the square symmetry of the
dot array.

\section{Conclusions}
\label{Conclusions}
We model the real-time excitation of a two-dimensional electron gas in a periodic square array
of quantum dots by solving the Liouville-von Neumann equation associated with a time-dependent
QEDFT Hamiltonian. The DFT component of the problem is non-linear in the density operator,
and the electron density, and relies on a recently proposed energy functional describing
the interaction of the electrons with randomly polarized cavity photons,
that has been extended to a two-dimensional electron gas in an external homogeneous magnetic field.

This QEDFT approach shows
that the spectra for the dipole-, the quadrupole-, the monopole-, and the rotational excitations are
all red-shifted by the polarization of the electron charge by the cavity photons.
The red-shifting of the excitation modes is in accordance with earlier many-body calculations
for configuration interacting electrons and cavity photons in small systems using truncated
but large Fock-spaces. The influence of the cavity photons on the excitation spectra of the
system can be compared for consistency to the changes to the orbital and spin magnetization
of the system regarding the dependence on the electron number and the external magnetic flux.
The calculations have moreover given us an insight into the strong Coulomb exchange forces in
the relatively short period square lattice of quantum dots when one electron resides in each dot
leading to a small blue-shift of the excitation spectra. In addition to the red-shift
of the excitation modes caused by the electron-photon interaction we notice how the
interaction increases the oscillator strength of the upper dipole peak, but reduces the
strength of the quadrupole peaks. These effects can be understood in terms of the electron
polarizability.

The QEDFT approach employs an exchange and correlation energy functional based on the adiabatic-connection fluctuation-dissipation theorem and concepts derived from the analysis of van der Waals interaction in physics and quantum chemistry. Instead of directly utilizing the electron-photon coupling or the Maxwell equations, this method represents a ``photon-free'' approach, but focusing solely on their resultant effects. Therefore questions regarding the number of photons, photon replicas, and
resonant transitions may thus not have obvious answers. Nonetheless, the presented calculations
show that, on the one hand this functional can reproduce results found by pure many-body approaches,
and on the other hand, predicts measurable effects where the more involved methods are difficult
to apply.

\begin{acknowledgments}
This work was financially supported by the Research Fund of the University
of Iceland, and the Icelandic Infrastructure Fund. The computations were performed on resources
provided by the Icelandic High Performance Computing Center at the University of Iceland.
V.\ Mughnetsyan and V.G.\ acknowledge support by the Armenian State Committee
of Science (grant No 21SCG-1C012). V.\ Mughnetsyan acknowledges support by the Armenian State
Committee of Science (grant No 21T-1C247).
V.\ Moldoveanu acknowledges financial support from the Core Program of the National Institute
of Materials Physics, granted by the Romanian Ministry of Research, Innovation and Digitalization
under the Project PC2-PN23080202.
\end{acknowledgments}

%

%
\frenchspacing
\bibliographystyle{apsrev4-2}
%
%

%
%
%
\end{document}